\documentclass[preprint,aps, amsmath,amssymb]{revtex4-1}
\newcommand{\be}{\begin{equation}}
\newcommand{\ee}{\end{equation}}
\newcommand{\bea}{\begin{eqnarray}}
\newcommand{\eea}{\end{eqnarray}}
\newcommand{\ba}{\begin{array}}
\newcommand{\ea}{\end{array}}

\usepackage[utf8]{inputenc}
\usepackage{graphicx}
\usepackage{epstopdf}
\usepackage{amsmath}
\usepackage{slashed}
\usepackage{mathtools}
\usepackage{slashed}
\usepackage{array,multirow}

\begin{document}

\title{Dynamical spin effects in the holographic light-front wavefunctions of light pseudoscalar mesons}

\author{Mohammad Ahmady}
\email{mahmady@mta.ca}
\affiliation{\small Department of Physics, Mount Allison University, \mbox{Sackville, New Brunswick, Canada, E4L 1E6}}

\author{Chandan Mondal}
\email{mondal@impcas.ac.cn}
\affiliation{Institute of Modern Physics, Chinese Academy of Sciences, Lanzhou-730000, China}

\author{Ruben Sandapen}
\email{ruben.sandapen@acadiau.ca, rsandapen@mta.ca}
\affiliation{\small Department of Physics, Acadia University, Wolfville, Nova Scotia, Canada, B4P 2R6}
\affiliation{\small Department of Physics, Mount Allison University, \mbox{Sackville, New Brunswick, Canada, E4L 1E6}}

\begin{abstract} 
We quantify the importance of dynamical spin effects in the holographic light-front wavefunctions of the pion, kaon, $\eta$ and $\eta^\prime$. Using a universal AdS/QCD scale and constituent quark masses, we find that such effects  are maximal in the pion where they lead to an excellent simultaneous description of a wide range of data: the decay constant, charge radius, spacelike EM and transition form factors, as well as, after QCD evolution, both the parton distribution function and the parton distribution amplitude data from Fermilab. These dynamical spin effects lead up to a $30\%$ chance of finding the valence quark and antiquark with aligned spins in the pion. The situation is very different for the kaon, where a simultaneous description of the available data (decay constant, radius and spacelike EM form factor) prefer no dynamical spin effects at all. The situation is less clear for the $\eta$ and $\eta^\prime$: while their radiative decay widths data are consistent with dynamical spin effects only in $\eta^\prime$, the data on their spacelike transition form factors clearly favor maximal dynamical spin effects in both mesons. 
\end{abstract}

\maketitle

\section{Introduction}
\label{Sec:Introduction}

A remarkable equation in hadronic physics is the holographic Schr\"odinger equation for mesons \cite{deTeramond:2005su,Brodsky:2006uqa,deTeramond:2008ht,Brodsky:2014yha}, 
 \begin{equation}
			\left(-\frac{\mathrm{d}^2}{\mathrm{d}\zeta^2}-\frac{1-4L^2}{4\zeta^2} + U_{\mathrm{eff}}(\zeta) \right) \phi(\zeta)=M^2 \phi(\zeta) \;,
	\label{hSE}
	\end{equation}
which is derived within a semiclassical approximation of light-front QCD, where quantum loops and quark masses are neglected. The holographic variable
\begin{equation}
	\mathbf{\zeta} = \sqrt{x\bar{x}} b \hspace{1cm} \hspace{1cm}  \hspace{1cm} (\bar{x} \equiv 1-x	)\;,
	\end{equation}
where $b$ is the transverse separation of the quark and antiquark and $x$ is the light-front momentum fraction carried by the quark, maps onto the fifth dimension, $z$, of anti$-$de Sitter (AdS) space so that Eq. (\ref{hSE}) also describes the propagation of weakly coupled string modes in a modified AdS space. The confining QCD potential, $U_{\mathrm{eff}}$, is then determined by the form of the dilaton field, $\varphi(z)$, which distorts the pure AdS geometry. Specifically, we have \cite{Brodsky:2014yha}
	\begin{equation}
	U_{\mathrm{eff}}(\zeta)= \frac{1}{2} \varphi^{\prime \prime}(z) + \frac{1}{4} \varphi^{\prime}(z)^2 + \frac{2J-3}{2 z} \varphi^{\prime}(z) \;.
	\label{dilaton-potential}
	\end{equation}	
A remarkable feature in light-front holographic QCD is that the form of the confinement potential is uniquely determined \cite{Brodsky:2013ar} to be that of a harmonic oscillator, i.e.  $U_{\mathrm{eff}}=\kappa^4 \zeta^2$ where $\kappa$ is the fundamental AdS/QCD scale. To recover this harmonic potential, the dilaton field has to be quadratic, i.e. $\varphi(z)=\kappa^2 z^2$ so that Eq. (\ref{dilaton-potential}) then implies that 
	\begin{equation}
		U_{\mathrm{eff}}(\zeta)=\kappa^4 \zeta^2 + 2 \kappa^2 (J-1)	
	\label{hUeff}	
	\end{equation}
	where $J=L+S$. Solving the holographic Schr\"odinger equation with the confining potential given by Eq. (\ref{hUeff}), yields the meson mass spectrum:
\begin{equation}
 	M^2= 4\kappa^2 \left(n+L +\frac{S}{2}\right)\;,
 	\label{mass-Regge}
 \end{equation}
and the wavefunctions
 \begin{equation}
 	\phi_{nL}(\zeta)= \kappa^{1+L} \sqrt{\frac{2 n !}{(n+L)!}} \zeta^{1/2+L} \exp{\left(\frac{-\kappa^2 \zeta^2}{2}\right)}  ~ L_n^L(\kappa^2 \zeta^2)\;.
 \label{phi-zeta}
 \end{equation}
The first nontrivial prediction is that the lightest bound state, with quantum numbers $n=L=S=0$, is massless: $M^2=0$.  This state is naturally identified with the pion since spontaneous chiral symmetry breaking in massless QCD implies a massless pion. However, here the massless pion does not result from spontaneous chiral symmetry breaking but is a consequence of the unique form of the holographic confining potential which is itself constrained by the de Alfaro, Furbini and Furlan (dAFF) \cite{deAlfaro:1976vlx} conformal symmetry breaking mechanism. The second prediction is that the meson masses lie on universal Regge trajectories, as is experimentally observed, with slopes determined by the AdS/QCD scale $\kappa$. This allows $\kappa$ to be extracted from spectroscopic data.  For vector mesons, $\kappa=540$ MeV reproduces the observed Regge slopes \cite{Brodsky:2014yha} and a similar value is used to successfully predict diffractive $\rho$ \cite{Forshaw:2012im} and $\phi$ electroproduction \cite{Ahmady:2016ujw}. For pseudoscalar mesons, a somewhat higher value, $\kappa=590$ MeV is required \cite{Brodsky:2014yha}. Note that this does not necessarily suggest that $\kappa=590$ MeV must be used for the pion because the latter, being very light, does not lie on Regge trajectory. This fact was previously highlighted in \cite{Vega:2008te} and indeed vastly different values of $\kappa$ for the pion have been used in the literature: \cite{Brodsky:2007hb,Brodsky:2011xx,Bacchetta:2017vzh,Vega:2009zb,Branz:2010ub,Swarnkar:2015osa}. In Ref. \cite{Ahmady:2016ufq}, we have shown that it might not be necessary to use a distinct value of $\kappa$ for the pion, provided that dynamical spin effects are taken into account. Indeed, we obtained very good agreement with the data when using $\kappa=523$ MeV, a value which itself results from a simultaneous fit to the Regge slopes of mesons and baryons \cite{Brodsky:2016rvj}, and accurately predicts the strong coupling to five-loop accuracy \cite{Deur:2016opc}.  We shall generate all predictions with $\kappa=(523 \pm 24)$ MeV \cite{Brodsky:2016rvj} in this paper. This value of $\kappa$ with the quoted uncertainty is what we refer to as the universal AdS/QCD scale.

 The holographic light-front Schr\"odinger Equation only gives the transverse part of the meson light-front wavefunction. The complete wavefunction is given by \cite{Brodsky:2014yha}
\begin{equation}
	\Psi (x,\zeta,\varphi)=\frac{\phi(\zeta)}{\sqrt{2\pi \zeta}} X(x) e^{iL\varphi} \;,
	\end{equation}
where $X(x)=\sqrt{x\bar{x}}$ as obtained by a precise mapping of the EM form factor in AdS and in physical spacetime \cite{Brodsky:2008pf}. The normalized holographic light-front wavefunction for a ground state meson is then given by
 \begin{equation}
 	\Psi (x,\zeta^2) = \frac{\kappa}{\sqrt{\pi}} \sqrt{x \bar{x}}  \exp{ \left[ -{ \kappa^2 \zeta^2  \over 2} \right] } \;.
\label{pionhwf} 
\end{equation}
Going beyond the semiclassical approximation, one can account for 
nonvanishing quark masses. This is carried out in Ref. \cite{Brodsky:2014yha}, yielding
\begin{equation}
\Psi (x,\zeta^2) = \mathcal{N} \sqrt{x \bar{x}}  \exp{ \left[ -\frac{\kappa^2 \zeta^2}{2} \right] } 
 \exp{ \left[ - {m_{f}^2 \bar{x} + m_{\bar{f}}^2 x \over 2 \kappa^2 x \bar{x} } \right]}
\label{pion-hwf-quark-masses}
\end{equation}
where $\mathcal{N}$ is a normalization constant which is fixed by requiring that 
 \begin{equation}
 	\int \mathrm{d}^2 \mathbf{b} \mathrm{d} x |\Psi(x,\zeta^2)|^2 = 1 \;. 
 	\label{norm}
 \end{equation}
 In Eq. \eqref{pion-hwf-quark-masses} and throughout this paper, $f=(q,s)$ with $q=(u,d)$, denoting the lighter up and down quarks. Note that Eq. \eqref{pion-hwf-quark-masses} can be recovered by adding a perturbation mass term to the effective potential of the holographic Schr\"odinger Equation,
 \begin{equation}
 	U_{\mathrm{eff}} \to U_{\mathrm{eff}} + \frac{m_f^2}{x} + \frac{m^2_{\bar{f}}}{\bar{x}} \;.
 \end{equation}
 This leads to a shift in the predicted meson masses,
\begin{equation}
	M^2 = M_0^{2} + \Delta M^2
\end{equation}
where $M_0^{2}$ stands for the meson mass squared in the limit of massless quarks [i.e. given by Eq. (\eqref{mass-Regge})] and the mass shift is given by \cite{Brodsky:2014yha}
\begin{equation}
	\Delta M^2 = \left\langle \Psi \left|\frac{m^2_f}{x} + \frac{m^2_{\bar{f}}}{\bar{x}}\right|\Psi \right \rangle \;.
\label{mass-shift}
\end{equation}
In Ref. \cite{Brodsky:2014yha}, the modified holographic wavefunction, given by Eq. \eqref{pion-hwf-quark-masses}, is used to compute $\Delta M^2$. Given that for the pion and kaon, $\Delta M^2=M^2_{\pi,K}$ (since $M^{2}_{0(\pi,K)}=0$), Ref. \cite{Brodsky:2014yha} uses these equalities to fix the quark masses $m_{q}$ ($q=u/d$) and $m_s$. With $\kappa=540$ MeV, this procedure yields $m_{q}=46$ MeV and $m_s=357$ MeV. 


As pointed out in Ref. \cite{Ahmady:2016ufq}, there is an implicit assumption within the semiclassical approximation of light-front QCD that the spins of the quarks decouple from their dynamics.  In other words, dynamical spin effects are neglected. Such effects were taken into account in Ref. \cite{Ahmady:2016ufq} for the pion and subsequently in Ref. \cite{Chang:2016ouf} for both the pion and the kaon, and furthermore to predict the decay constants of light and heavy-light mesons in Ref. \cite{Chang:2018aut}. In this paper, we differ from and hopefully complement the analysis of Ref. \cite{Chang:2016ouf} in several ways. First, we use a different ansatz for the spin structure of a pseudoscalar meson. Secondly, we simultaneously predict  the decay constant and the charge radius and EM form factor, while the latter two observables are not considered in Ref. \cite{Chang:2016ouf}. A simultaneous description of the decay constant and the radius is interesting since the radius quantifies the departure of the meson from a pointlike particle (and thus is sensitive to long-distance physics), while the decay constant is sensitive to the wavefunction at zero transverse separation, i.e. to short-distance physics. Therefore, a successful simultaneous description of the radius and decay constant is a stringent test on any model of the meson wavefunction. Third, we shall also investigate the importance of dynamical spin effects in the $\eta$ and $\eta^\prime$ mesons. In Ref. \cite{Momeni:2017moz}, dynamical spin effects are taken into account in the kaon holographic wavefunction in order to predict the decay $B \to K \mu^+ \mu^-$, although the focus is not on whether such effects are actually required by the kaon data. One of our goals in this paper is to investigate whether the data show evidence for dynamical spin effects in pseudoscalar mesons heavier than the pion. Finally, we shall also go beyond the previous pion analysis by two of us in Ref. \cite{Ahmady:2016ufq} by taking into account QCD evolution that allows us to compare our predictions for the holographic pion parton distribution function (PDF) and parton distribution amplitude (PDA) with the data from the E615 \cite{Conway:1989fs} and E791 \cite{Aitala:2000hb} Collaborations respectively. In particular, we shall perform fits to the  E615 data in order to determine the low hadronic scale (which was left unspecified in Ref. \cite{Ahmady:2016ufq}) at which our predictions are valid.

\section{Dynamical spin effects}
\label{Spin}
Assuming that the spin structure of a pseudoscalar meson results from the pointlike coupling of a pseudoscalar particle to a $q\bar{q}$ pair and that all bound state effects are captured by the holographic light-front wavefunction, we can write the spin-improved wavefunction as 
\begin{equation}
	\Psi^{P}_{h \bar{h}}(x,\mathbf{k}) = \Psi(x, \mathbf{k}) S^P_{h \bar{h}} (x, \mathbf{k})
	\label{spin-improved-wf}
	\end{equation}
where 
\begin{equation}
	S^P_{h \bar{h}} (x, \mathbf{k})= \frac{\bar{u}_{h}(x,\mathbf{k})}{\sqrt{\bar{x}}} \left[A \frac{M_P}{2P^+} \gamma^+ \gamma^5 + B  \gamma^5 \right] \frac{v_{\bar{h}}(x,\mathbf{k})}{\sqrt{x}} 
	\label{spin-structure} 
	\end{equation}
and $\Psi(x, \mathbf{k})$ is the holographic meson light-front wavefunction. $A$ and $B$ are dimensionless, arbitrary constants. Note that this spin structure is a special case of the more general spin structure ($A(P\cdot \gamma) \gamma^5 + B M_P \gamma^5$ \cite{Leutwyler:1973mu,Dziembowski:1987zp,Ji:1990rd}) where, as in \cite{Chang:2016ouf,Li:2017mlw,Leitner:2010nx,Carbonell:1998rj}, we retain only the $\gamma^+ \gamma^5$ component of the $A$ term. We emphasize that this can be done since it does not break the boost invariance of the light-front wavefunction. Our motivation for not keeping the $\gamma^- \gamma^5$ term is purely phenomenological: retaining that term does not lead to agreement with the data.

Equation (\eqref{spin-structure}) differs in two ways from that used in Ref. \cite{Chang:2016ouf,Chang:2018aut} where  the invariant  mass of the $q\bar{q}$ pair replaces the physical meson mass and no factor of $\sqrt{x\bar{x}}$ appears in the denominator. As we shall show below, our ansatz has the advantage of reducing to the original holographic wavefunction when $B \to 0$, i.e. it possesses a limit in which the spins of the quarks decouple from their dynamics. On the other hand, if $B \to \infty$, we say that dynamical spin effects are maximal. For these reasons, it is appropriate to call $B$ the dynamical spin parameter. Note that the limits $B \to 0$ or $B \to \infty$ (numerically $B \gg 1$) formally correspond to selecting either the axial-vector or the pseudoscalar spin structure in Eq. \eqref{spin-structure}.  However, we prefer to interpret them as the limits of vanishing or maximal dynamical spin effects since this is more reminiscent of their physical meaning in our approach. Note that such an interpretation would not be correct if the axial-vector structure is chosen as $\slashed{P} \gamma_5$, as in Ref. \cite{Momeni:2017moz}, where taking $B=0$ does not imply vanishing dynamical spin effects.

We now simplify Eq. \eqref{spin-structure} using the light-front spinors and matrices given in Ref. \cite{Lepage:1980fj}. Inserting
\begin{equation}
	\frac{\bar{u}_{h}}{\sqrt{x}} \gamma^5 \frac{v_{\bar{h}}}{\sqrt{\bar{x}}}=\frac{h(xm_f+\bar{x}m_{\bar{f}})}{x\bar{x}} \delta_{h,-\bar{h}}- \frac{ke^{-ih\theta_k}}{x\bar{x}}\delta_{h,\bar{h}} 
\label{gamma5}
\end{equation} 
and
\begin{equation}
	\frac{\bar{u}_{h}}{\sqrt{x}} \gamma^+\gamma^5 \frac{v_{\bar{h}}}{\sqrt{\bar{x}}}=h 2P^+ \delta_{h,-\bar{h}}\;.
\label{pgamma5}
\end{equation} 
into  Eq. \eqref{spin-structure}, we obtain
	\begin{eqnarray}
		S_{h,\bar{h}}^P(x, \mathbf{k})=\frac{1}{x\bar{x}} \left[ (AM_P x\bar{x} + B(xm_f + \bar{x} m_{\bar{f}}) )h \delta_{h,-\bar{h}}  - B k e^{-ih\theta_k} \delta_{h,\bar{h}}	\right]
	\label{explicit-S}
	\end{eqnarray}
	where $\mathbf{k}=ke^{i\theta_k}$. It therefore follows that
	\begin{eqnarray}
	 	\Psi_{h,\bar{h}}^{P}(x,\mathbf{k})= \left[ (AM_P x\bar{x} + B (xm_f + \bar{x} m_{\bar{f}}) ) h\delta_{h,-\bar{h}}  - B    k e^{-ih\theta_k}\delta_{h,\bar{h}}	\right] \frac{\Psi (x,\mathbf{k}^2)}{x\bar{x}} \;.
	 \label{spin-improved-wfn-k}
	 \end{eqnarray}
	After a two-dimensional Fourier transform of Eq. \eqref{spin-improved-wfn-k}, we obtain
	 \begin{eqnarray}
	 	\Psi_{h,\bar{h}}^{P}(x,\mathbf{b})= \left[ (AM_P x\bar{x} + B(xm_f + \bar{x} m_{\bar{f}})  ) h\delta_{h,-\bar{h}} + i B h   \partial_b \delta_{h,\bar{h}}	\right] \frac{\Psi (x,\zeta^2)}{x\bar{x}}
	 	\label{spin-improved-wfn-b}
	 \end{eqnarray}
 where $\Psi(x,\zeta^2)$ is the two dimensional Fourier transform of the holographic meson wavefunction given by Eq. \eqref{pion-hwf-quark-masses}. As we mentioned above, our wavefunction possesses a nondynamical spin limit. In fact, with $A=\kappa/\sqrt{2 \pi} M_P$ and $B=0$, we recover exactly the normalized original holographic wavefunction, Eq. \eqref{pion-hwf-quark-masses}:
 \begin{equation}
 	\Psi_{h\bar{h}}^{P} (x,\zeta^2) = \frac{\kappa}{\sqrt{\pi}} \sqrt{x \bar{x}}  \exp{ \left[ -{ \kappa^2 \zeta^2  \over 2} \right] } \exp{ \left[ - {m_{f}^2 \bar{x} + m_{\bar{f}}^2 x \over 2 \kappa^2 x \bar{x} } \right]} \times \frac{1}{\sqrt{2}} h\delta_{h,-\bar{h}} 
\end{equation}
 with a nondynamical spin wavefunction.

 Our spin-improved wavefunction is normalized using
	\begin{equation}
 	\int \mathrm{d}^2 \mathbf{b} \mathrm{d} x |\Psi^{P}(x,\mathbf{b}^2)|^2 = 1 
 	\label{norm-spin}
 \end{equation}	
 where now
 \begin{equation}
 	|\Psi^{P}(x,\mathbf{b}^2)|^2 =\sum_{h,\bar{h}} |\Psi_{h,\bar{h}}^{P}(x,\mathbf{b}^2)|^2 	\;.
 	\label{sum-notation}
 	\end{equation}
 Note that our normalization condition, Eq. \eqref{norm-spin}, implies that we exclude any contributions due to higher Fock states. The normalization condition allows us to set $A=1$.  Then, in addition to the quark masses, there remains only one free parameter: the dynamical spin parameter $B$. We emphasize that with the introduction of dynamical spin effects, the effective quark masses can no longer be fixed using Eq. \eqref{mass-shift}.  They are thus, \textit{\`a priori}, free parameters. In this paper, we shall show that constituentlike quark masses, lead to the best agreement with the data. 
 

\section{Decay constants and radiative decay widths}
 \label{Decay constants}
The decay constant of a pseudoscalar meson is defined as 
\begin{equation}
	\langle 0 | J^P_{\mu 5} | P \rangle =  f_{P} P_\mu \;.
\label{fM-def}
\end{equation}
For the charged pion and kaon, 
\begin{equation}
	J^{P}_{\mu 5}=\bar{u} \gamma_\mu \gamma_5 q   
\end{equation}
where $q=d~\mathrm{and}~s$, respectively. Taking $\mu=+$ and expanding the left-hand side of Eq. \eqref{fM-def},  we obtain 
\begin{equation}
	\langle 0 | \bar{u} \gamma^+ \gamma^5  d | P \rangle=\sqrt{4 \pi N_c} \sum_{h, \bar{h}} \int \frac{\mathrm{d}^2 \mathbf{k}}{16\pi^3} \mathrm{d} x \Psi_{h,\bar{h}}^{P}(x,\mathbf{k}) \left \{\frac{\bar{v}_{\bar{h}}}{\sqrt{\bar{x}}} (\gamma^+ \gamma^5 )\frac{u_{{h}}}{\sqrt{{x}}}\right \} \;,
	\label{me:gamma+gamma5}
	\end{equation}
so that using our spin-improved holographic wavefunction, Eq. \eqref{spin-improved-wfn-b}, we find
\begin{equation}
	f_{P}(m_f,m_{\bar{f}},M_P,B)= 2 \sqrt{\frac{N_c}{\pi}}  \int \mathrm{d} x   [x\bar{x} M_P+ B (x m_f+\bar{x}m_{\bar{f}})]  \left.\frac{\Psi (x,\zeta)}{x\bar{x}}\right|_{\zeta=0}	
\label{decayconstant}
\end{equation}
Our predictions for the decay constants of the charged pion and kaon are shown in Table \ref{tab:decay-constants}. It can be seen that maximal dynamical spin effects ($B \gg 1$) are favored for the pion, while for the kaon the opposite is true: the measured kaon decay constant prefers no dynamical spin effects ($B=0$).

\begin{table}
  \centering
  \begin{tabular}{|c|c|c|c|}
    \hline
$P$ &B&$f^{\mathrm{Th.}}_{P}$ [MeV]&$f^{\mathrm{Exp.}}_{P}$[MeV]\\
\hline
 &$0$ & $162 \pm 8$&  \\
\cline{2-3}
$\pi^{\pm}$&$1$ & $138 \pm 5$& $130 \pm 0.04 \pm 0.2$ \\
\cline{2-3}
 &$\gg 1$ & $135 \pm 6$&  \\\hline        
 &$0$ & $156 \pm 8$ &  \\
 \cline{2-3}
 $K^{\pm}$&$1$& $ 142\pm 7$ & $156 \pm 0.5$\\
 \cline{2-3}
 &$\gg 1$& $135 \pm 6$ &  \\
\hline 
         
\end{tabular}
  \caption{Our predictions for the decay constants of the charged pion and kaon, compared to the measured values from Particle Data Group (PDG) \cite{Patrignani:2016xqp}. The theory uncertainties result from the uncertainties in the constituent quark masses and the AdS/QCD scale: $[m_q,m_s]=([330,500] \pm 30)$ MeV and $\kappa=(523 \pm 24)$ MeV.}
  \label{tab:decay-constants}
\end{table}

For the $\eta$ and $\eta^\prime$, we need to account for quantum mechanical mixing. In the SU(3) octet-singlet basis,
\begin{eqnarray}
	|\eta_1\rangle &=&\frac{1}{\sqrt{3}}\left(|u\bar{u}\rangle + |d\bar{d}\rangle + |s\bar{s}\rangle \right)\\ \nonumber
	|\eta_8\rangle &=&\frac{1}{\sqrt{6}}\left(|u\bar{u}\rangle + |d\bar{d}\rangle -2 |s\bar{s}\rangle \right) \;, 
\end{eqnarray}
the physical $\eta$ and $\eta^\prime$ states are given by 
\begin{eqnarray}
\label{mix}	
	|\eta\rangle &=& \cos \theta |\eta_8 \rangle - \sin \theta |\eta_1 \rangle \\ \nonumber
	|\eta^\prime\rangle &=&\sin \theta |\eta_8 \rangle + \cos \theta |\eta_1 \rangle 
\end{eqnarray}
where $\theta$ is the mixing angle. Here we use $\theta=-14.1^\circ \pm 2.8^\circ$, as determined by the lattice simulation by the RBC-UKQCD Collaboration \cite{Christ:2010dd},  which is somewhat intermediate between the measurements of the KLOE Collaboration \cite{Ambrosino:2006gk}: $\theta=-15.0 \pm 0.7$ with a gluonium content in $\eta^\prime$, and the average value $\theta=-11.5^\circ$ obtained using the quadratic and linear form of the Gell-Mann-Okubo mass formula. Inverting Eq. \eqref{mix} and using the light-front Schr\"odinger equation, $H_{\mathrm{LF}}|P\rangle = M^2 |P\rangle$ \cite{Brodsky:1997de}, we can express the masses of $\eta_1$ and $\eta_8$ in terms of the physical $\eta$ and $\eta^{\prime}$ masses as follows:
\begin{eqnarray}
M_{\eta_1}^2&=&\sin^2 \theta M_{\eta}^2 + \cos^2 \theta M^2_{\eta^\prime}\\ \nonumber
M_{\eta_8}^2&=&\cos^2 \theta M_{\eta}^2 + \sin^2 \theta M^2_{\eta^\prime}	\;.
\end{eqnarray}
The decay constants of $\pi^0$, $\eta_1$ and $\eta_8$ can then be computed using the axial-vector currents
\begin{equation}
	J^{\pi^0}_{\mu 5}=\frac{1}{\sqrt{2}}(\bar{u} \gamma_\mu \gamma_5 u - \bar{d} \gamma_\mu \gamma_5 d) \;, 
\end{equation}
\begin{equation}
	J^{\eta_8}_{\mu 5}=\frac{1}{\sqrt{6}}(\bar{u} \gamma_\mu \gamma_5 u + \bar{d} \gamma_\mu \gamma_5 d -2 \bar{s} \gamma_\mu \gamma_5 s) \;, 
\end{equation}
and
\begin{equation}
	J^{\eta_1}_{\mu 5}=\frac{1}{\sqrt{3}}(\bar{u} \gamma_\mu \gamma_5 u + \bar{d} \gamma_\mu \gamma_5 d + \bar{s} \gamma_\mu \gamma_5 s)  
\end{equation}
respectively. Assuming isospin symmetry, this leads to 
\begin{equation}
	f_{\pi^0}= f_P(m_{q},M_{\pi^0},B)=f_{\pi^{\pm}} \;,
\end{equation}
\begin{equation}
	f_{\eta_1}= \frac{1}{3}[2f_P(m_{q},M_{\eta_1},B_{q})+ f_P(m_s,M_{\eta_1},B_s)]
\end{equation}
and
\begin{equation}
	f_{\eta_8}= \frac{1}{3}[f_P(m_{q},M_{\eta_8},B_{q})+ f_P(m_s,M_{\eta_8},B_s)] 
\end{equation}
where $f_P(m_{q/s},M_P,B_{q/s})$ is given by Eq. \eqref{decayconstant}. Notice that, based on our findings for the charged pion and kaon decay constants, we have allowed the dynamical spin parameter $B$ to differ in the nonstrange and the strange sectors of the $\eta$ and $\eta^\prime$. We now use the Alder-Bell-Jackiw (ABJ) anomaly relations \cite{Adler:1969gk,Bell:1969ts} to compute the photon-meson transition form factor at zero momentum transfer as follows \cite{Choi:2017zxn}:
\begin{equation}
	F_{\pi \gamma}^{\mathrm{ABJ}}(0)=\frac{1}{2\sqrt{2}\pi^2f_{\pi^0}} \;,
\end{equation}

\begin{equation}
	F_{\eta \gamma}^{\mathrm{ABJ}}(0)=\frac{1}{2\sqrt{6}\pi^2}\left(\frac{1}{f_{\eta_8}}\cos \theta - \frac{2\sqrt{2}}{f_{\eta_1}} \sin \theta \right) \;,
\end{equation}
and
\begin{equation}
	F_{\eta^\prime \gamma}^{\mathrm{ABJ}}(0)=\frac{1}{2\sqrt{6}\pi^2}\left(\frac{1}{f_{\eta_8}}\sin \theta + \frac{2\sqrt{2}}{f_{\eta_1}} \cos \theta \right) \;,
\end{equation}
so that we can predict the radiative decay widths using
\begin{equation}
	\Gamma_{P \to \gamma \gamma}= \frac{\pi}{4}\alpha^2_{\mathrm{EM}} M_P^3 |F_{P\gamma}(0)|^2 
\end{equation}
where $P=\pi^0,\eta,\eta^\prime$. Our results are shown in Table \ref{tab:2photon}. We see that the pion and $\eta^\prime$ data are consistent with dynamical spin effects which can even be maximal. For the $\eta^\prime$, this statement is still true even if we restrict the dynamical spin effects only to its nonstrange $q\bar{q}$ sector. On the other hand, for the $\eta$, the measured radiative decay width prefers no dynamical spin effects at all. We shall come back to the $\eta{-}\eta^\prime$ system when  we consider their spacelike transition form factors in Section \ref{TFFs}.

\begin{table}
    \begin{tabular}{|c|c|c|c|c|}\hline
    $P$ & $B_{q}$ &$B_s$& $\Gamma^{\mathrm{Th.}}_{P \to 2 \gamma}$[KeV] & $\Gamma^{\mathrm{Exp.}}_{P \to 2 \gamma}$[KeV] \\ \cline{1-5}
\multirow{2}{*}
    {$\pi^0$} 
    & 0 & -& $(5.62 \pm 0.57)\times 10^{-3}$ &  \\
    \cline{2-4}
    & $1$ & -&$(7.72 \pm 0.62)  \times 10^{-3}$  & $(7.82 \pm 0.22)\times 10^{-3}$ \\ 
    \cline{2-4}    
    & $\gg 1$ & -&$(8.13 \pm 0.68) \times 10^{-3}$  & \\ \hline
\multirow{2}{*}
    {$\eta$} 
    & 0 & 0& $0.542 \pm 0.082$  & \\
    \cline{2-4}
    & $1$ & $0$ & $ 0.600 \pm 0.056$ &  \\     
    \cline{2-4}
    & $1$ & $1$ & $ 0.622 \pm 0.055$ & $0.516 \pm 0.018$ \\     
    \cline{2-4}
    & $\gg 1$ & 0 & $ 0.663 \pm 0.061$ &  \\  
    \cline{2-4}   
    & $\gg 1$ &$\gg 1$& $0.710 \pm 0.059$ & \\ \hline
\multirow{2}{*}
    {$\eta^\prime$} 
    & 0 & 0&$3.51 \pm 0.48$  &  \\
    \cline{2-4}    
    & $1$ & $0$ & $3.88  \pm 0.49$ & \\     
    \cline{2-4}    
    & $1$ & $1$ & $3.94  \pm 0.49$ & $4.28 \pm 0.19$\\     
    \cline{2-4}    
    & $\gg 1$ & 0 & $4.51 \pm 0.56$ & \\ 
    \cline{2-4}
   & $\gg 1$ & $\gg 1$& $4.73 \pm 0.57$ & \\ \hline
\end{tabular}
 \caption{Our predictions (computed using the ABJ anomaly relations) for the radiative decay width, $\Gamma_{P \to 2\gamma}$, compared to the measured PDG average values \cite{Patrignani:2016xqp}. The theory uncertainties result from the uncertainties in the constituent quark masses, the AdS/QCD scale and, additionally, the mixing angle for the $\eta$ and $\eta^\prime$: $[m_q,m_s]=([330,500] \pm 30)$ MeV, $\kappa=(523 \pm 24)$ MeV and $\theta=14.1^\circ \pm 2.8^\circ$.}
  \label{tab:2photon}
  \end{table}

\section{Charge radii and EM form factors}
\label{radius-decay-EMFF}
\begin{figure}[htbp]
\centering 
\includegraphics[width=8cm,height=8cm]{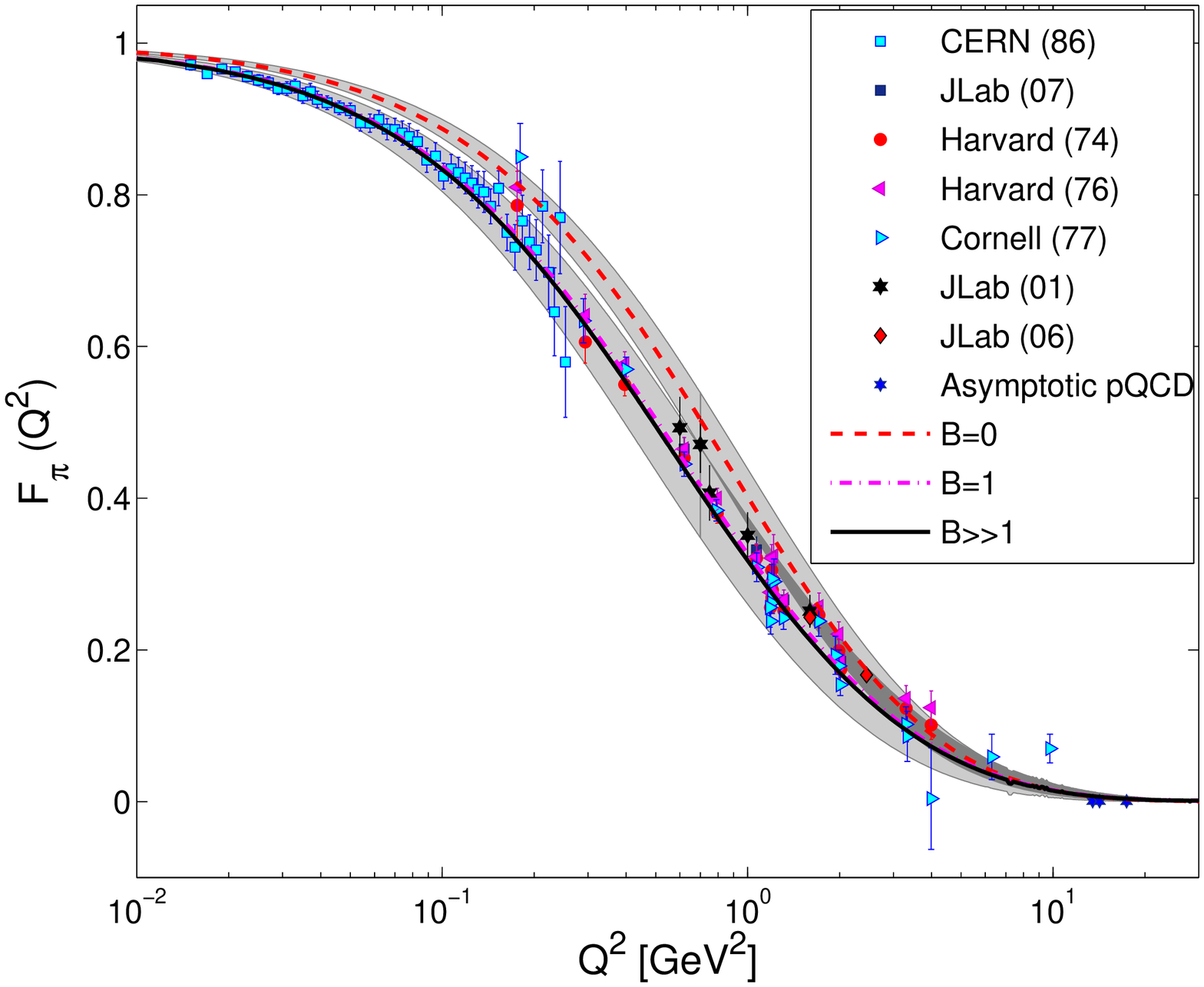}\includegraphics[width=8cm,height=8cm]{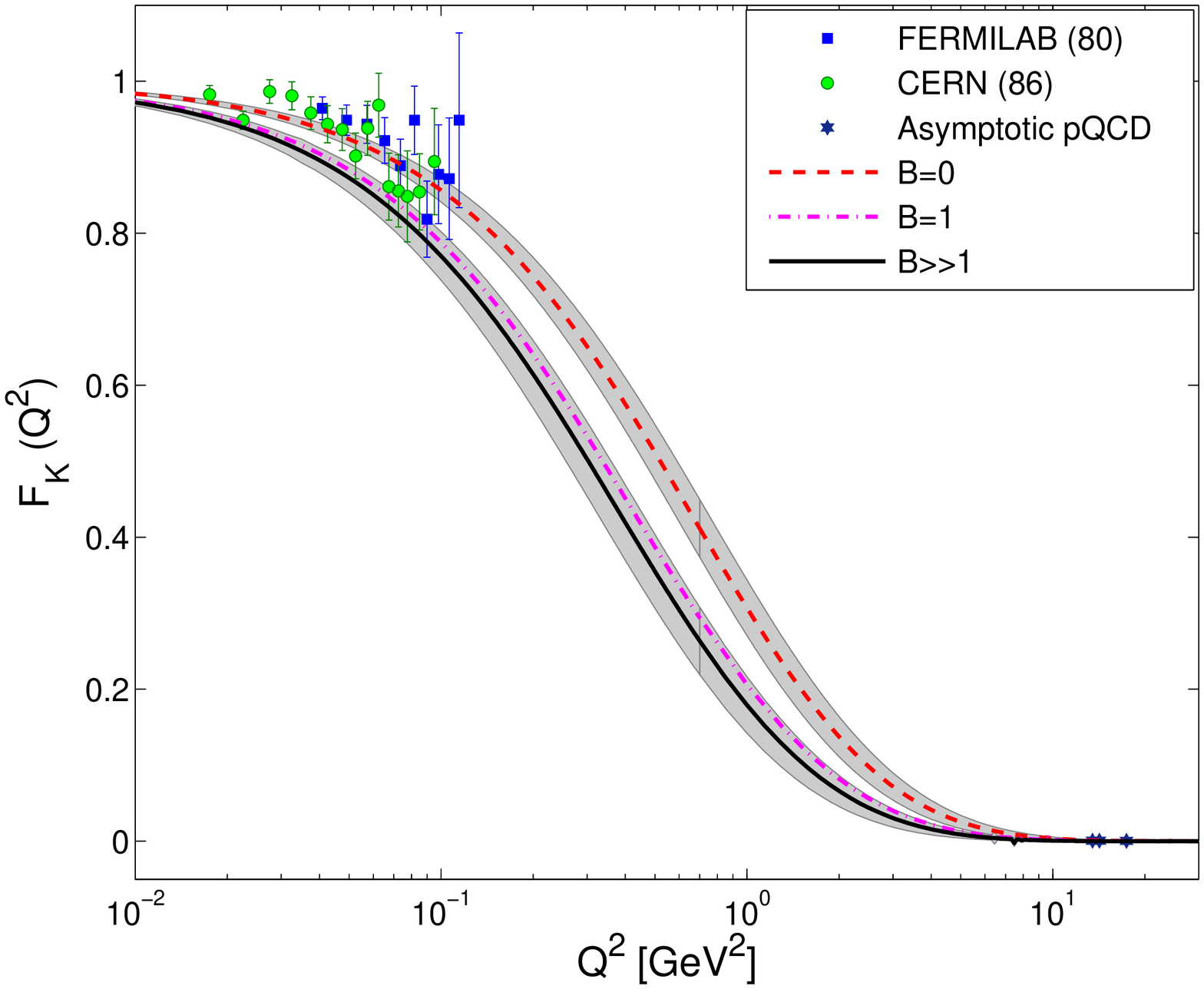}
\caption{Our predictions for the pion (left) and kaon (right) EM form factors. Dashed red curves: $B=0$. Orange dotted-dashed curves: $B=1$. Solid black curves: $B \gg 1$. The grey bands for the $B=0$ and $B \gg 1$ curves indicate the theory uncertainty resulting from the uncertainties in the constituent quark masses and the AdS/QCD scale: $[m_q,m_s]=([330,500] \pm 30)$ MeV and $\kappa=(523 \pm 24)$ MeV. Data from Refs. \cite{Amendolia:1986wj,Bebek:1974,Bebek:1976,Bebek:1978,Volmer:2001,Horn:2006tm,Pedlar:2005sj,Seth:2012nn} for the pion form factor and from Refs. \cite{Dally:1980dj,Amendolia:1986ui,Seth:2012nn,Pedlar:2005sj} for the kaon form factor.} 
\label{Fig:EMFF}
\end{figure}

The root-mean-square pion radius can be computed using \cite{Brodsky:2007hb}
\begin{equation}
	\sqrt{\langle r_{P}^2 \rangle} = \left[\frac{3}{2} \int \mathrm{d} x \mathrm{d}^2 \mathbf{b} [b (1-x)]^2 |\Psi^{P}(x,\mathbf{b})|^2 \right]^{1/2} 
	\label{radius}
\end{equation}
where $|\Psi^{P}(x,\mathbf{b})|^2$ is given by Eq. \eqref{sum-notation}. The EM form factor is defined as
\begin{equation}
	\langle  P (p^{\prime})| J_{\text{EM}}^\mu (0) | P (p) \rangle = 2 (p + p^{\prime})^\mu F_{P}(Q^2)
\end{equation}
where $p^{\prime}=p+q$, $Q^2=-q^2$ and the EM current $J_{\mathrm{EM}}^\mu(z)=\sum_f e_f \bar{\Psi} (z) \gamma^\mu \Psi(z)$ with $f=\bar{d},u$ and $e_{\bar{d},u}=1/3,2/3$. The EM form factor can be expressed in terms of the meson light-front wavefunction using the Drell-Yan-West formula as follows \cite{Drell:1969km,West:1970av}:
\begin{equation}
	F_{P}(Q^2)= 2 \pi \int \mathrm{d} x \mathrm{d} b ~ b ~ J_{0}[(1-x)  b Q] ~ |\Psi^{P}(x,\textbf{b})|^2 
\label{DYW}
\end{equation}
where $|\Psi^{P}(x,\textbf{b})|^2$ is given by Eq. \eqref{sum-notation}. Note that Eq. \eqref{DYW} implies that $F_{\pi}(0)=1$ if the meson light-front wavefunction is normalized according to Eq. \eqref{norm-spin} and that the slope of the EM form factor at $Q^2=0$ is related to the mean radius of the pion given by Eq. \eqref{radius} via 
\begin{equation}
	\langle r_{P}^2 \rangle = -\frac{6}{F_{P}(0)} \left . \frac{\mathrm{d} F_{P}}{\mathrm{d} Q^2} \right|_{Q^2=0} \;.
\end{equation}
Our predictions for the charge radii are compared to the PDG values in Table \ref{tab:radii}, and our predictions for the EM form factors are compared to the available data in Fig. \ref{Fig:EMFF}.  
\begin{table}
  \centering
  \begin{tabular}{|c|c|c|c|}
    \hline
$P$ &$B$&$\sqrt{\langle r_{P}^2 \rangle}_{\mathrm{Th.}}$ [fm]&$\sqrt{\langle r_{P}^2 \rangle}_{\mathrm{Exp.}}$[fm]\\
\hline
$\pi^{\pm}$ &$0$ & $0.543 \pm 0.022$&  \\
\cline{2-3}
&$1$ & $ 0.673\pm 0.034 $& $0.672 \pm 0.008$ \\
\cline{2-3}
&$\gg 1$ & $0.684 \pm 0.035$&  \\
\hline
$K^{\pm}$ &$0$ & $0.615 \pm 0.038$ & \\
\cline{2-3}
 &$1$& $ 0.778 \pm 0.065$ & $0.560 \pm 0.031$  \\
\cline{2-3}
 &$\gg 1$& $0.815 \pm 0.070$ &  \\
\hline 
         
\end{tabular}
  \caption{Our predictions for the radii of $\pi^{\pm}$ and $K^{\pm}$, compared to the measured values from PDG \cite{Patrignani:2016xqp}. The theory uncertainties are due to the uncertainties in the constituent quark masses and in the AdS/QCD scale: $[m_q,m_s]=([330,500] \pm 30)$ MeV and $\kappa=(523 \pm 24)$ MeV.}
  \label{tab:radii}
\end{table}
It is clear from Table \ref{tab:radii} and Fig. \ref{Fig:EMFF} that the pion data favor maximal dynamical spin effects. These results were already found in Ref. \cite{Ahmady:2016ufq}, but we now include the theory uncertainties due to the AdS/QCD scale and quark masses to show that these uncertainties do not alter our previous conclusion. In particular, the agreement with the precise low $Q^2$ EM form factor data remains very impressive. On the other hand, we now find that the kaon data actually prefer no dynamical spin effects at all. The kaon data set is limited to the decay constant, charge radius and EM form factor but for the pion, both the PDF and the PDA have been measured at Fermilab, giving us further opportunity to investigate if those data also reveal maximal dynamical spin effects.

\section{Parton Distribution Function}
\label{PDF}

Our holographic valence pion parton distribution function is given by 
\begin{equation}
	f_{v}(x)=  \int \mathrm{d}^2 \mathbf{b}   |\Psi_{\pi} (x,\mathbf{b})|^2 
	\label{DA}
\end{equation}
where $\Psi_{\pi}(x,\mathbf{b})$ is given by Eq. \eqref{spin-improved-wfn-b}. This holographic PDF is valid at a low hadronic scale, $\mu_0$, which can be  determined from data \cite{Bacchetta:2017vzh}. At any scale, the PDF satisfies the normalization condition,
\begin{equation}
	\int \mathrm{d} x f_{v}(x,\mu)  =1 \;,
\end{equation}
so that for $\mu \to \infty$, Eq. \eqref{norm-spin} is recovered. We shall now fix $\mu_0$ by fitting, after QCD evolution, to the pion PDF data from the E615 experiment \cite{Conway:1989fs}. We evolve the PDF to $\mu=5$ GeV (relevant to the E615 experiment) using the evolution code for modified Dokshitzer-Gribov-Lipatov-Altarelli-Parisi (DGLAP) equations \cite{Zhu:1999ht,Zhu:1998hg,Zhu:2004xi,Zhu:1999jm}, although we notice that the higher twist corrections to the DGLAP equations are not important for the evolution of the valence quark distribution. 

\begin{figure}[htbp]
\centering 
\includegraphics[width=8cm,height=8cm]{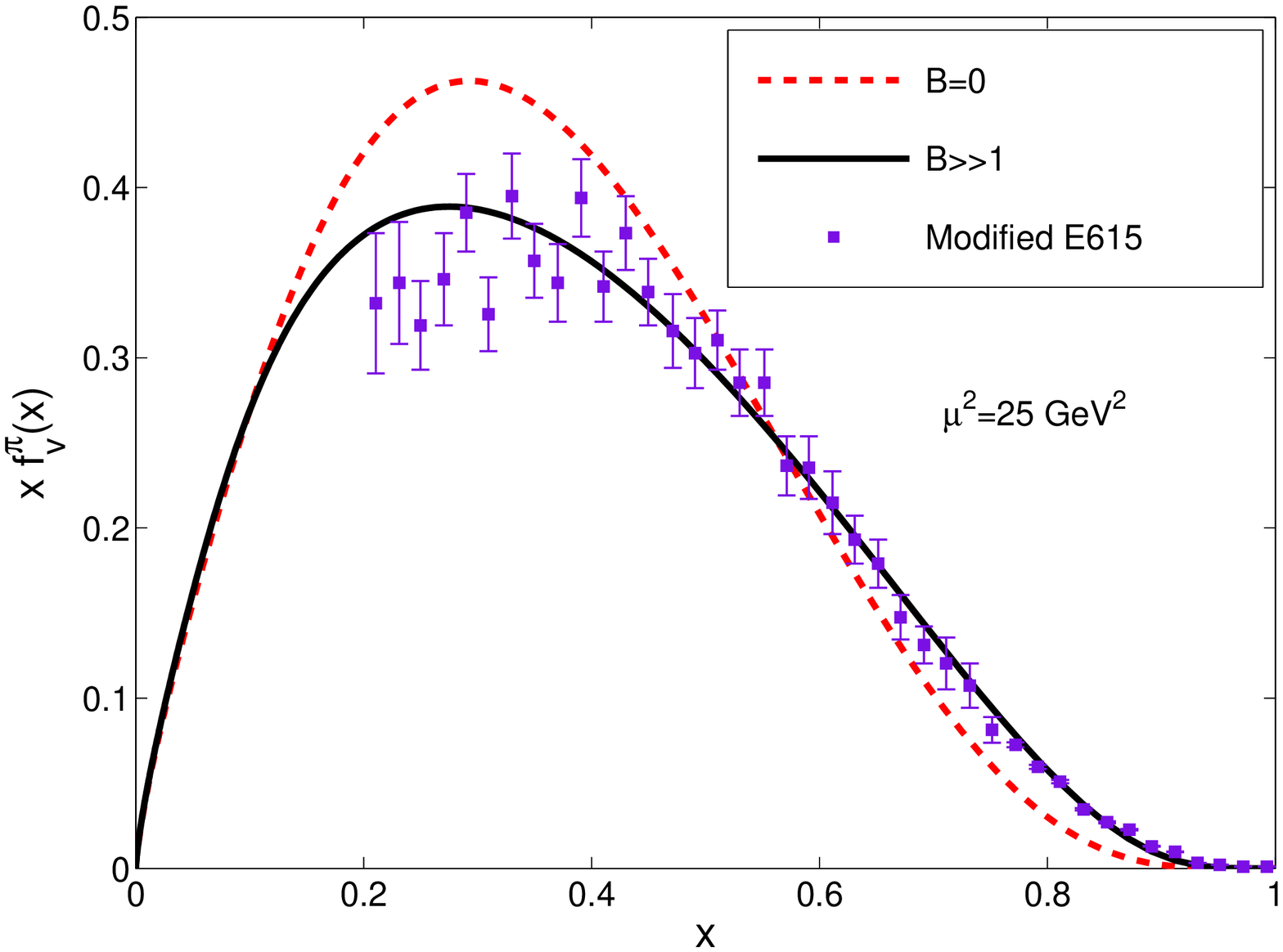}\includegraphics[width=8cm,height=8cm]{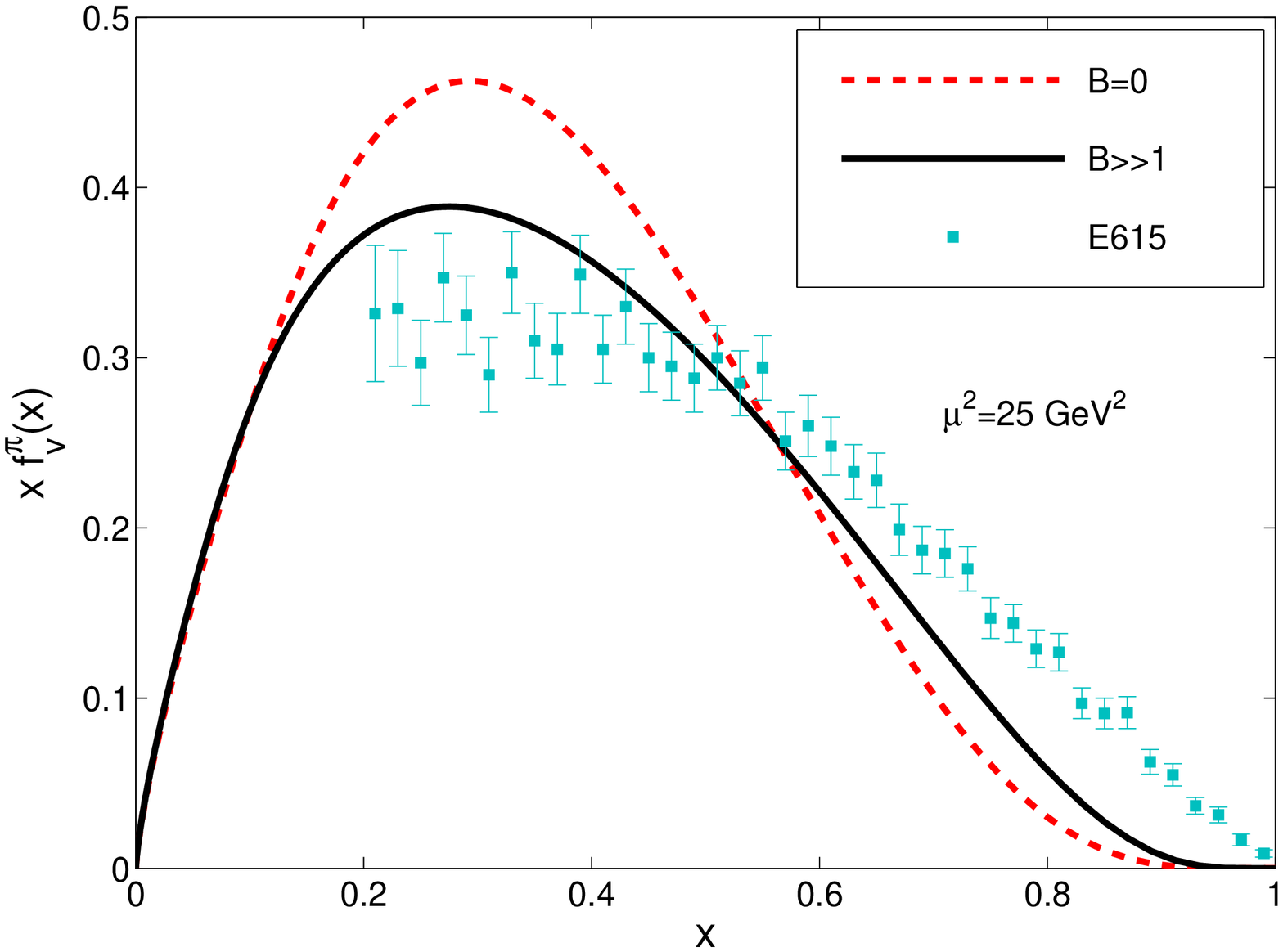}\caption{Left: Our fits to the original E615 data \cite{Conway:1989fs}. Right: Our fits to the modified E615 data \cite{Aicher:2010cb}. QCD evolution is carried out using DGLAP evolution \cite{Zhu:1999ht,Zhu:1998hg,Zhu:2004xi,Zhu:1999jm} with an initial scale $\mu_{0}=0.316$ GeV. Solid black curves: $B \gg 1$. Dashed red curves: $B=0$. The $B=1$ curves  almost coincide with the $B \gg 1$ curves and therefore are not shown here.} 
\label{Fig:PDF}
\end{figure}

In Fig. \ref{Fig:PDF}, we show our fits to the modified E615 data after being reanalyzed to take into account soft gluon resummation \cite{Aicher:2010cb}. We are able to fit the reanalyzed data using an initial scale $\mu_0=0.316$ GeV (similar to the initial scale used in \cite{Gutsche:2013zia,RuizArriola:2002bp}), with $\Lambda_{\mathrm{QCD}}=0.226$ GeV and no initial sea quarks or gluons. The fit is best achieved with $B \gg 1$. With $B=0$, we could not find another initial scale that yielded a better fit. Our fit to the data is very good, especially at large $x>0.6$, and is thus in agreement with the Dyson-Schwinger analysis of Ref. \cite{Chen:2016sno}. In a recent paper \cite{deTeramond:2018ecg}, the HLFHS Collaboration reports a successful fit to the original E615 data \cite{Conway:1989fs} in light-front holographic QCD with massless quarks and no dynamical spin effects. Instead, there is a nonzero probability that the pion consists of the higher Fock state $|q\bar{q}q\bar{q}\rangle$, as determined from the analysis of the timelike pion EM form factor in AdS/QCD \cite{Brodsky:2014yha}. The initial scale is taken to be equal to the transition scale between AdS/QCD and pQCD: $\mu_0=(1.1 \pm 0.2)$ GeV. Note that, while we go beyond the semiclassical approximation by accounting for nonzero quark masses and dynamical spin effects, we explicitly exclude higher Fock states in the pion via the normalization given by Eq. \eqref{norm-spin}. For comparison with the result in Ref. \cite{deTeramond:2018ecg}, we also show our attempt to fit to the original E615 data in Fig. \ref{Fig:PDF}. As can be seen, we are not successful, especially for large $x >0.6$ region of the data. 
  
  The behavior of the pion PDF at large $x$ is still an unresolved issue \cite{Holt:2010vj,deTeramond:2018ecg} and therefore we cannot make a definite statement about the importance of dynamical spin effects in the pion based solely on our fit to the reanalyzed E615 data. However, if future experiments confirm the reanalysis of the E615 data set, then it will add to the existing evidence (based on the decay constant, radius and EM form factor data) that dynamical spin effects are indeed maximal in the pion. In any case, the fits to the PDF data allow us to fix the scale of 
  our model predictions to be $\mu_0=0.316$ GeV. 
  
\section{Parton Distribution Amplitude}
\label{PDA}
Our holographic parton distribution amplitude for a pseudoscalar meson is given by 
\begin{equation}
	\varphi_{P} (x;m_f,m_{\bar{f}},B)=  \frac{2}{f_P} \sqrt{\frac{N_c}{\pi}} \int \mathrm{d}^2 \mathbf{b}  \{((x(1-x) M_P^2)+ B (xm_f + (1-x)m_{\bar{f}})M_P\} \frac{\Psi (x,\zeta)}{x(1-x)} 
	\label{DA}
\end{equation}
which we assume to be valid, like the pion PDF, at $\mu_{0}=0.316$  GeV. The normalization condition for the PDA at any scale is:
\begin{equation}
	\int \mathrm{d} x \varphi_P(x,\mu)=1
\end{equation}
where we have suppressed the dependence of the Distribution Amplitude (DA) on the quark masses and dynamical spin parameter. Note that for $\mu \to \infty$, Eq. \eqref{decayconstant} is recovered. The LO QCD evolution of the PDA is carried using the Efremov-Radyushkin-Brodsky-Lepage (ERBL) equations \cite{Efremov:1978rn,Efremov:1979qk,Lepage:1980fj}. In a Gegenbauer basis, we have \cite{RuizArriola:2002bp}
\begin{equation}
	\varphi(x,\mu)= 6x\bar{x} \sum_{n=0}^{\infty} C_n^{3/2}(2x-1) a_n(\mu)
\end{equation}
where
\begin{equation}
	a_n(\mu)=\frac{2}{3} \frac{2n+3}{(n+1)(n+2)} \left(\frac{\alpha_s(\mu)}{\alpha_s(\mu_{0})}\right)^{\gamma_n^{0}/2\beta_0} \int_0^1 \mathrm{d}x C_n^{3/2}(2x-1) \varphi(x,\mu_0) \;.
\label{an}
\end{equation}
with
\begin{equation}
	\gamma_n^{(0)}=-2C_F \left[3+ \frac{2}{(n+1)(n+2)}-4\sum_k^{n+1} \frac{1}{k} \right]
\end{equation}
and
\begin{equation}
	\beta_0=\frac{11}{3} C_A -\frac{2}{3}n_f \;.
\end{equation}
The strong running coupling in Eq. \eqref{an} is given by
\begin{equation}
	\alpha_s (\mu) =\frac{4\pi}{\beta_0\ln \left(\mu^2/\Lambda^2_{\mathrm{QCD}}\right)} 
\end{equation}
where we take $n_f=3$ as the number of active flavors and $\Lambda_{\mathrm{QCD}}=0.226$ GeV for the evolution of the PDF \cite{Gutsche:2013zia}. The moments of the DA are directly related to its Gegenbauer coffecients. For the first four moments, we have \cite{Choi:2007yu}
\begin{equation}
	\langle \xi_1 \rangle = \frac{3}{5} a_1 \;,
\end{equation}
\begin{equation}
	\langle \xi_2 \rangle = \frac{12}{35} a_2 + \frac{1}{5} \;,
\end{equation}
\begin{equation}
	\langle \xi_3 \rangle = \frac{9}{35} a_1 + \frac{4}{31} a_3\;,
\end{equation}
\begin{equation}
	\langle \xi_4 \rangle = \frac{3}{35} + \frac{8}{35} a_2 + \frac{8}{77} a_4 \;.
\end{equation}

\begin{figure}[htbp]
\centering 
\includegraphics[width=8cm,height=8cm]{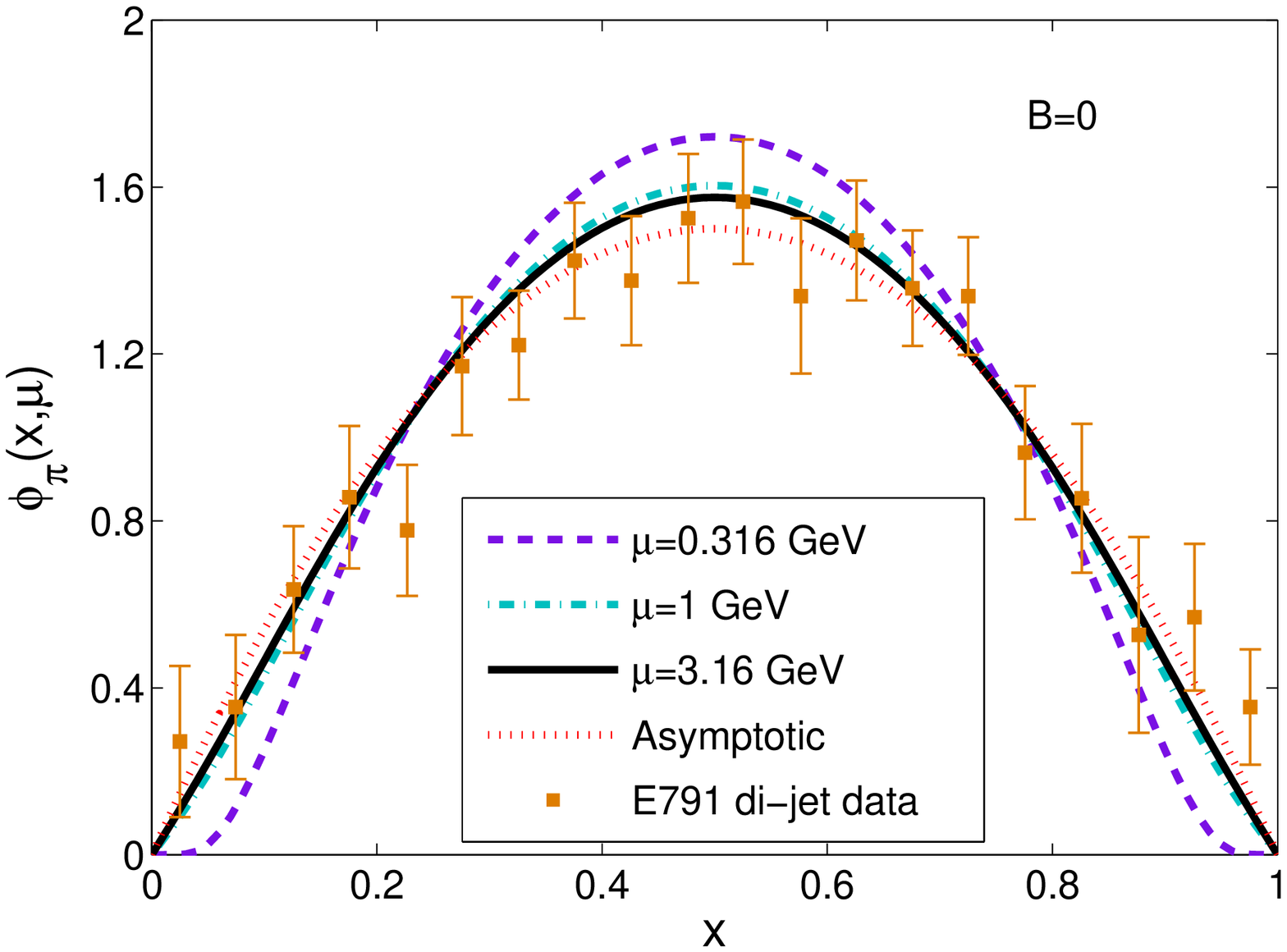}\includegraphics[width=8cm,height=8cm]{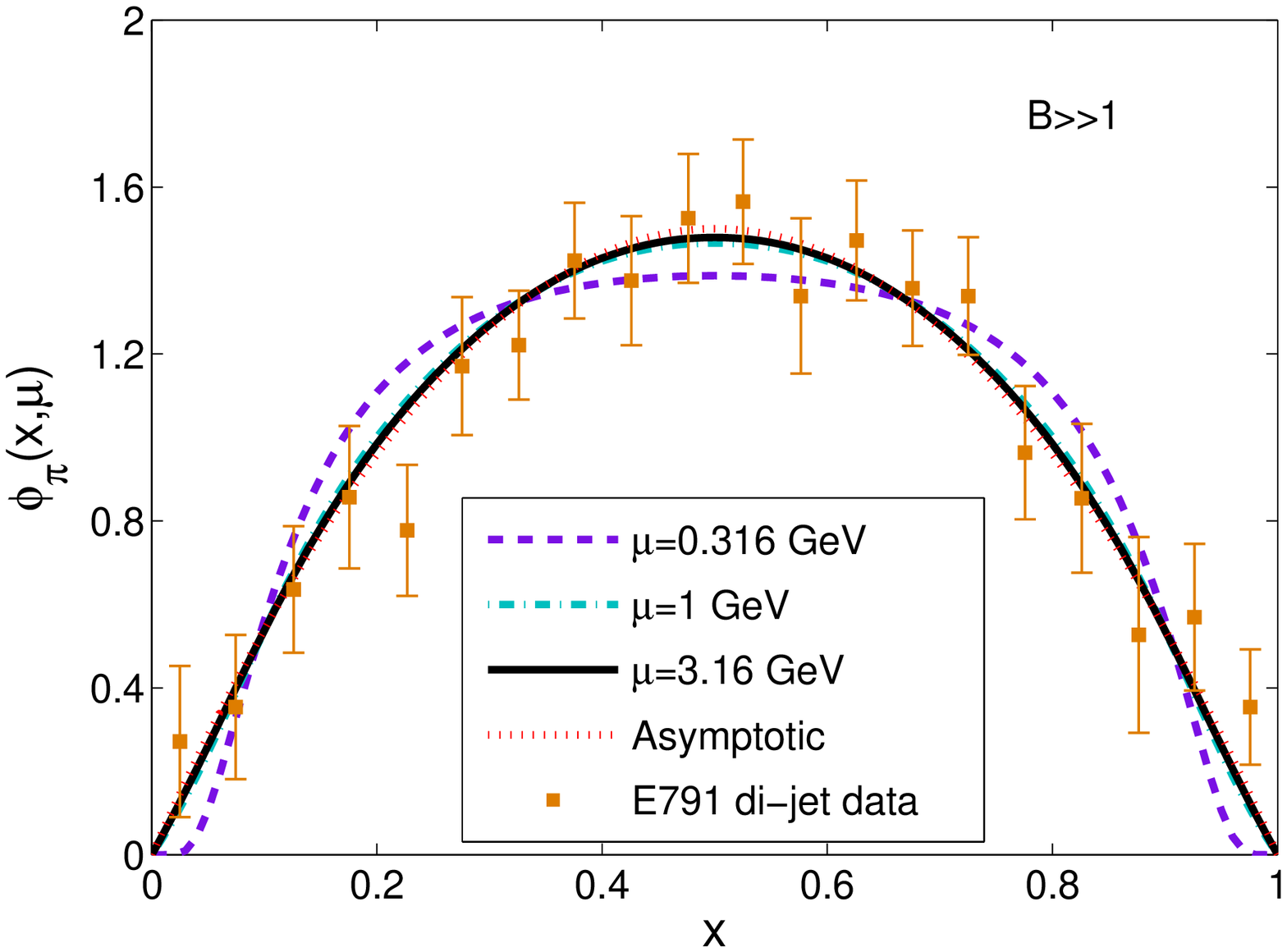}\caption{QCD evolution of our holographic pion DAs from $\mu_0=0.316$ GeV (dashed purple curves) to $\mu=1$ GeV (dotted-dashed cyan curves) and finally to $\mu=3.16$ GeV (solid black curves), which is the scale relevant to the E791 data \cite{Aitala:2000hb}. The asymptotic DAs are the dotted red curves.} 
\label{Fig:PionDA}
\end{figure}

Our predictions for the moments of the pion holographic DAs are compared to other theoretical predictions in Table \ref{tab:PionDAmoments}. There are two points worth noting regarding Table \ref{tab:PionDAmoments}. First, only the holographic DAs have their moments smaller than their asymptotic values and thus they evolve towards the latter from below. Second, at a given scale, dynamical spin effects bring the moments closer to their asymptotic values. This means that the spin-improved holographic DA evolves faster to the asymptotic DA than the original holographic DA. This faster evolution is confirmed in Fig. \ref{Fig:PionDA} where we show the evolution of both holographic DAs from $\mu_0=0.316$ GeV up to $\mu=3.16$ GeV which is the scale relevant to the E791 data \cite{Aitala:2000hb}. As can be seen, with dynamical spin effects, the holographic pion DA is almost identical to the asymptotic DA already at $\mu=1$ GeV. However, the E791 data are not precise enough to really discriminate between the two holographic DAs.

\begin{table}
  \centering
  \begin{tabular}{|c|c|c|c|c|}
    \hline
 Pion DA   & $\mu$ [GeV] &$\langle \xi_2 \rangle$ &$\langle \xi_4 \rangle$&$\langle x^{-1} \rangle$ \\
     \hline
     
     Asymptotic & $\infty$ & $0.200$ &$0.085$ & $3.00$ \\
    \hline 
    LF Holographic ($B=0$) &$1,2$  & $0.180,0.185$ &$0.067,0.071$ &$2.81,2.85 $\\
        \hline 
    LF Holographic ($B \gg 1$) &$1,2$  & $0.200,0.200$ &$0.085,0.085$&$2.93,2.95 $\\
    \hline
    LF Holographic \cite{Brodsky:2007hb} &$ \sim 1$  & $0.237$ &$0.114$&$4.0$\\
    \hline
         Platykurtic \cite{Stefanis:2014nla} & $2$ & $0.220^{+0.009}_{-0.006}$& $0.098^{+0.008}_{-0.005}$ & $3.13^{+0.14}_{-0.10}$\\
    \hline
    LF Quark Model \cite{Choi:2007yu} & $\sim 1$ & $0.24 [0.22]$ &$0.11 [0.09]$ & \\
    \hline
    Sum Rules \cite{Ball:2004ye} & $1$ & $0.24$ & $0.11$&\\
    \hline
       Renormalon model \cite{Agaev:2005rc}& $1$ & $0.28$ & $0.13$&\\
     \hline  
     Instanton  vacuum \cite{Petrov:1998kg,Nam:2006au}  & $1$ & $0.22,0.21$ & $0.10,0.09$&\\ 
          \hline    
          NLC Sum Rules \cite{Bakulev:2001pa} &$2$ &$0.248^{+0.016}_{-0.015}$  & $0.108^{+0.05}_{-0.03}$ &$3.16^{+0.09}_{-0.09}$ \\
    \hline
    Sum Rules\cite{Chernyak:1983ej}& $2$ & $0.343$ &$0.181$ &$4.25$\\
    
    \hline  
    Dyson-Schwinger[RL,DB]\cite{Chang:2013pq} & $2$ & $0.280,0.251$ &$0.151,0.128$ &$5.5,4.6$\\
     
    \hline
    Lattice \cite{Arthur:2010xf} &$2$& $0.28(1)(2)$& & \\
    \hline
   
      Lattice \cite{Braun:2015axa} & $2$ & $0.2361(41)(39)$ & &\\
     \hline
     Lattice \cite{Braun:2006dg} & $2$ & $0.27 \pm 0.04$ & &\\
     \hline
    \end{tabular}
  \caption{Our predictions for the first two nonvanishing moments and the inverse moment of the pion holographic DA, compared to other theoretical predictions.}
  \label{tab:PionDAmoments}
\end{table}

\begin{figure}[htbp]
\centering 
\includegraphics[width=8cm,height=8cm]{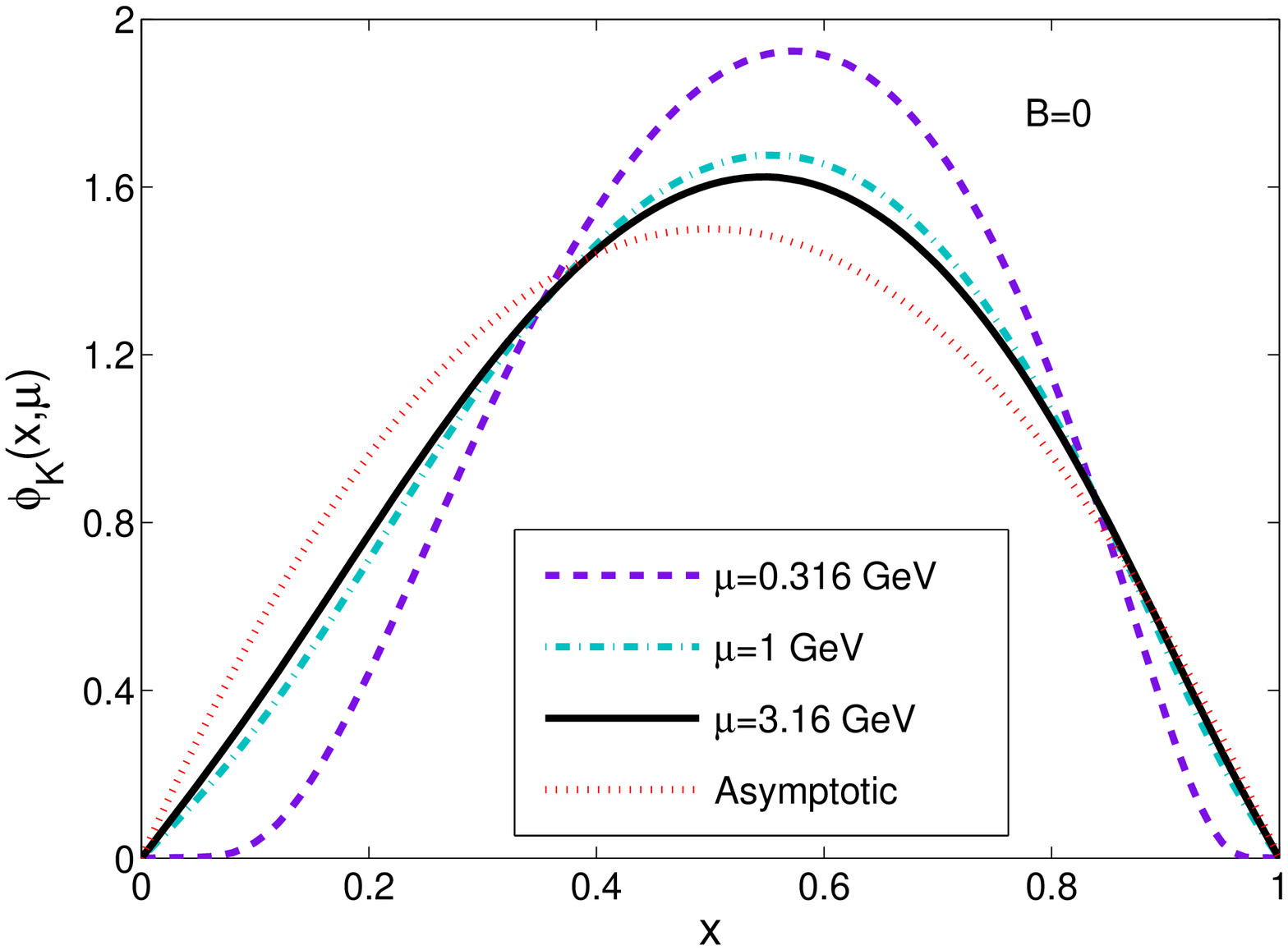}\includegraphics[width=8cm,height=8cm]{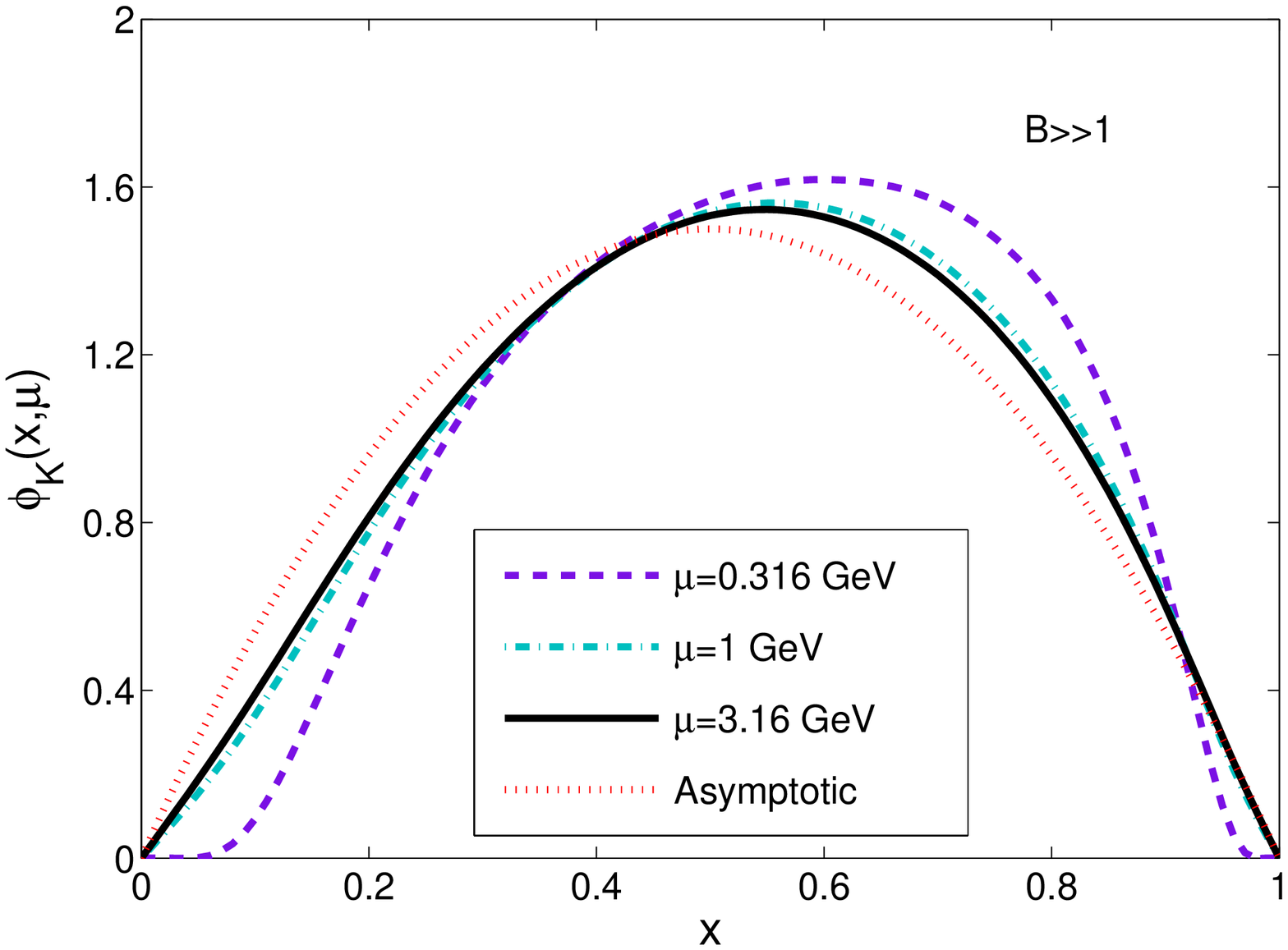}\caption{QCD evolution of our kaon holographic DAs from $\mu_0=0.316$ GeV (dashed purple curves) to $\mu=1$ GeV (dotted-dashed cyan curves) and up to $\mu=3.16$ GeV (solid black curves). The asymptotic DAs are the dotted red curves.} 
\label{Fig:KaonDA}
\end{figure}

Our predictions for the moments of the kaon holographic DAs  are compared to other theoretical predictions in Table \ref{tab:KDAmoments}. In this case, the predicted even and inverse moments are lower than their asymptotic values while the predicted odd moments are greater than zero (which is their asymptotic value). Dynamical spin has opposite effects on the predicted even or odd moments: the even moments are brought closer to their asymptotic values while the odd moments deviate further from their asymptotic value. As for the pion, the predicted even moments are lower than the other theoretical predictions shown here. No obvious trend emerges from a comparison of the odd moments of the various theoretical models. In Fig. \ref{Fig:KaonDA}, we show the evolution of the two holographic kaon DAs. As for the pion, dynamical spin effects enhance the evolution of the DA towards the asymptotic DA but, unlike the pion, the spin-improved holographic DA is still distinct from the asymptotic DA at $\mu=1$ GeV or even at $\mu=3.16$ GeV.

\begin{table}
  \centering
  \begin{tabular}{|c|c|c|c|c|c|c|}
    \hline
  Kaon DA   & $\mu$ [GeV] &$\langle \xi_1 \rangle$ & $\langle \xi_2 \rangle$&$\langle \xi_3 \rangle$ &$\langle \xi_4 \rangle$&$\langle x^{-1} \rangle$ \\
     \hline
     
     Asymptotic & $\infty$ &0&$0.200$ &0&$0.085$ & $3.00$ \\
    \hline 
    Holographic ($B=0$) &$1,2$  &$0.055,0.047$ &$0.175,0.180$ &$0.021,0.018$ & $0.062,0.067$ &$2.55,2.62$\\
        \hline 
    Holographic ($B \gg 1$) &$1,2$&$0.094,0.081$ &  $0.194,0.195$ &$0.039,0.034$&$0.080,0.081$ & $2.60,2.66$\\
   \hline
   
   Lattice \cite{Arthur:2010xf} &$2$&$0.036(2)$&$0.26(2)$&&& \\
   
    \hline
   LF Quark Model \cite{Choi:2007yu} &$\sim 1$&$0.06[0.08]$&$0.21[0.19]$&$0.03[0.04]$&$0.09[0.08]$& \\
   \hline
   Sum Rules \cite{Ball:2006wn} &$1$&$0.036$&$0.286$&$0.015$&$0.143$&$3.57$\\
      \hline
   Dyson-Schwinger[RL,DB] \cite{Shi:2014uwa}& $2$ & $0.11,0.040$ & $0.24,0.23$ & $0.064, 0.021$ & $0.12, 0.11$ & \\
   \hline
   Instanton vacuum \cite{Nam:2006au}&$1$&$0.057$&$0.182$&$0.023$&$0.070$&\\
   \hline
  \end{tabular}
  \caption{Our predictions for the first four and inverse moments of the kaon holographic DA compared to other theoretical predictions.}
  \label{tab:KDAmoments}
\end{table}

\section{Transition Form Factors}
\label{TFFs}
\begin{figure}[htbp]
\centering 
\includegraphics[width=8cm,height=8cm]{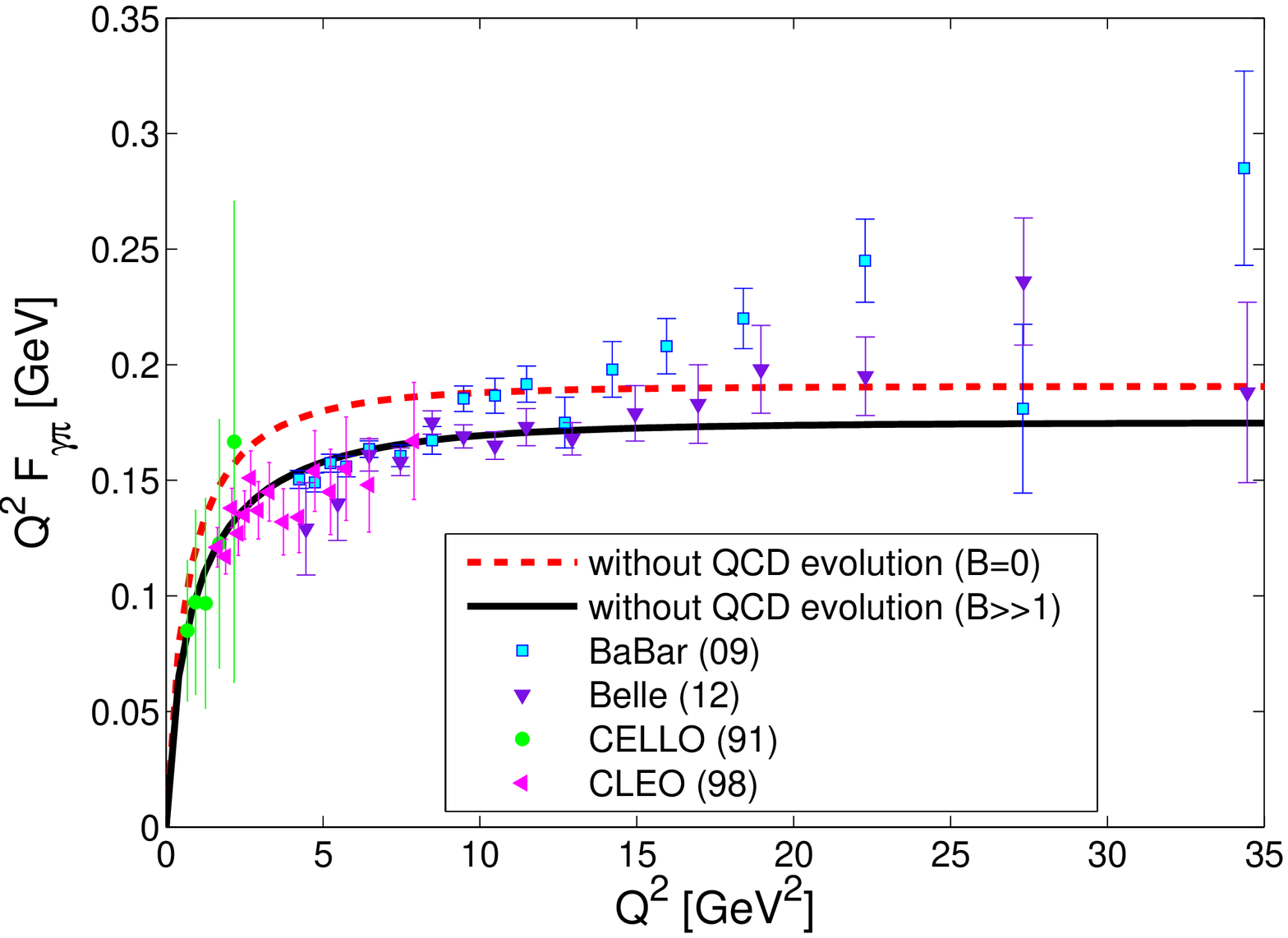}
\includegraphics[width=8cm,height=8cm]{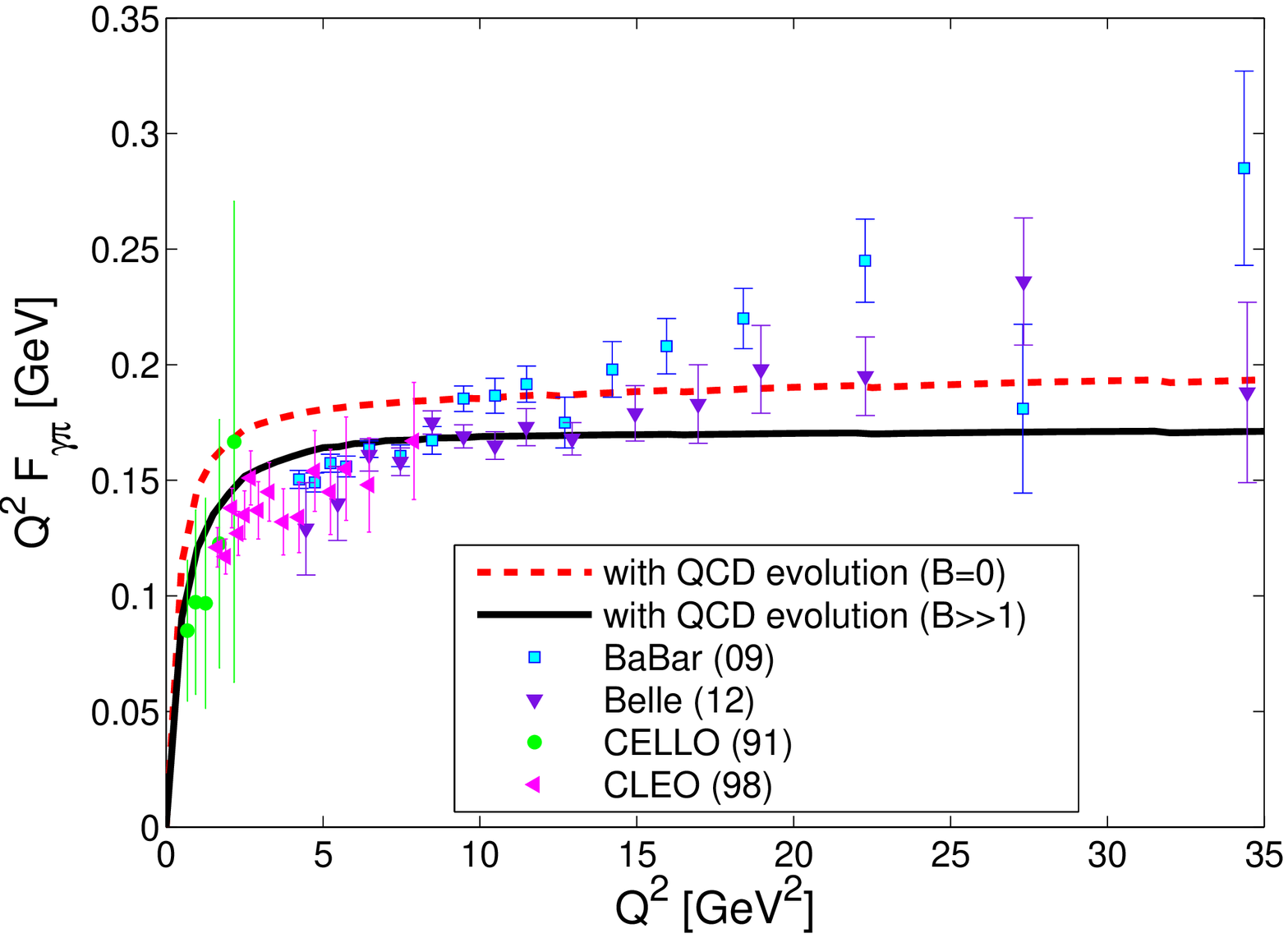}
\caption{Our predictions for the photon-to-pion transition form factor without QCD evolution (left) and with QCD evolution (right). Dashed red curves: $B=0$. Solid black curves: $B \gg 1$. The $B=1$ curves almost coincide with the $B \gg 1$ curves and are thus not shown here. The data are from the CELLO \cite{Behrend:1990sr}, CLEO \cite{Gronberg:1997fj}, BaBar \cite{Aubert:2009mc} and Belle \cite{Uehara:2012ag} Collaborations.} 
\label{fig:TFF}
\end{figure}

The photon-meson transition form factor is directly related to the inverse moment of the meson's PDA. For the pion, we have \cite{Choi:2017zxn}
\begin{equation}
	F_{\gamma \pi} (Q^2)= \left(\frac{\hat{e}_{u}^2-\hat{e}_{d}^2}{\sqrt{2}}\right) I(Q^2;m_{q},M_\pi,B)\;,
	\label{TFF}
\end{equation}
where \cite{Lepage:1980fj}
\begin{equation}
	I(Q^2;m_q,M_\pi,B)= 2 \int_0^1 \frac{\varphi_P(x,xQ; m_q,M_\pi,B)}{x Q^2} 
\end{equation}
and $\hat{e}_{u,d}=[2/3,-1/3]$. The DA, given by Eq. \eqref{DA}, is evaluated at a scale $\mu=xQ$ \cite{Brodsky:2011yv}. In Ref. \cite{Ahmady:2016ufq}, we already showed that dynamical spin effects in the pion lead to an excellent description of the photon-pion transition form factor data, although they cannot account for the so-called BaBar anomaly: the 2009 BaBar data set which exhibits strong scaling violations.  However, QCD evolution was not taken into account in Ref. \cite{Ahmady:2016ufq}. We now account for QCD evolution effects and, as can be seen in Fig. \ref{fig:TFF}, they confirm our previous conclusion: the data prefer maximal dynamical spin effects and we still cannot account for the BaBar anomaly. 

We now turn to the $\eta$ and $\eta^\prime$. Taking into account mixing, their transition form factors are given by 
\begin{eqnarray}
	F_{\eta \gamma} &=& \cos \theta F_{\eta_8 \gamma} - \sin \theta F_{\eta_1 \gamma} \\ \nonumber
	F_{\eta^\prime \gamma} &=& \sin \theta F_{\eta_8 \gamma} + \cos \theta F_{\eta_1 \gamma} \\ \nonumber
\end{eqnarray}	
where
\begin{equation}
	F_{\eta_1 \gamma}(Q^2)= \left(\frac{\hat{e}_{u}^2+\hat{e}_{d}^2}{\sqrt{3}}\right) I(Q^2;m_{q},M_{\eta_1},B_{q}) + \frac{\hat{e}_s^2}{\sqrt{3}} I(Q^2;m_s, M_{\eta_1},B_s)
\end{equation}
and
\begin{equation}
	F_{\eta_8\gamma}(Q^2)= \left(\frac{\hat{e}_{u}^2+\hat{e}_{d}^2}{\sqrt{6}}\right) I(Q^2;m_{q},M_{\eta_8},B_{q})-2 \frac{\hat{e}_s^2}{\sqrt{6}} I(Q^2;m_s, M_{\eta_8},B_s) \;.
\end{equation}
In the SU(3) chiral limit $\{m_q,m_s,M_P\} \to 0$, the transition form factors of the 3 mesons differ only by a constant factor: $F_{\eta_1 \gamma}(Q^2)=(2\sqrt{2}/\sqrt{3}) F_{\pi \gamma}(Q^2)$ and  $F_{\eta_8 \gamma}(Q^2)=(1/\sqrt{3}) F_{\pi \gamma}(Q^2)$.  These relations are indeed used in Ref. \cite{Brodsky:2011xx} to compute $F_{\eta \gamma}$ and $F_{\eta^\prime \gamma}$ in holographic light-front QCD with massless quarks. Perhaps more surprisingly, the same relations are used to compute $F_{\eta \gamma}$ and $F_{\eta^\prime \gamma}$ in Ref. \cite{Swarnkar:2015osa}, despite the fact that Ref. \cite{Swarnkar:2015osa} uses constituent quark masses where SU(3) flavor symmetry is broken: $[m_{q},m_s]=[330,500]$ MeV. No dynamical spin effects are investigated in Refs. \cite{Brodsky:2011xx,Swarnkar:2015osa}. 

\begin{figure}[htbp]
\centering 
\includegraphics[width=8cm,height=8cm]{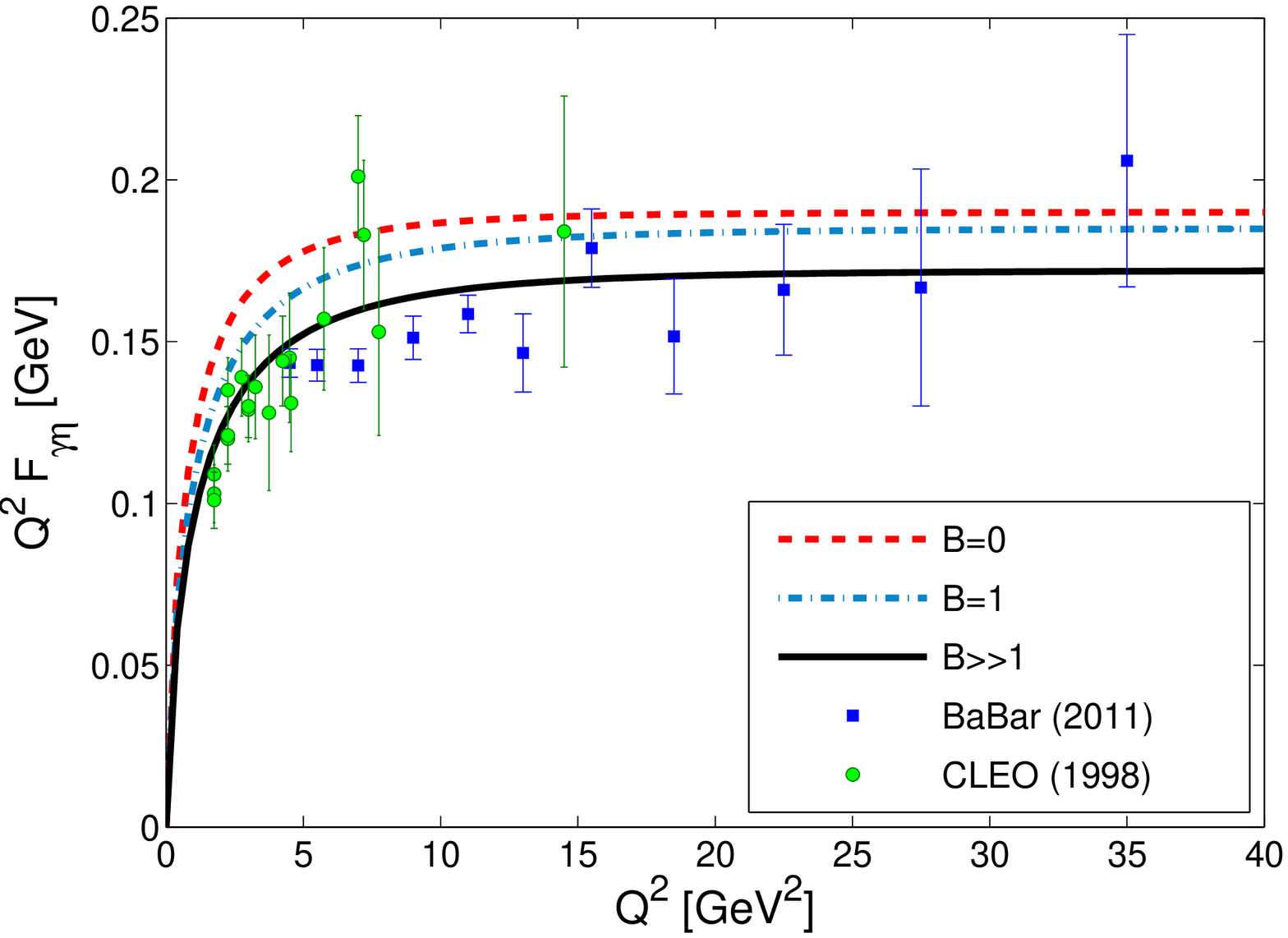}
\includegraphics[width=8cm,height=8cm]{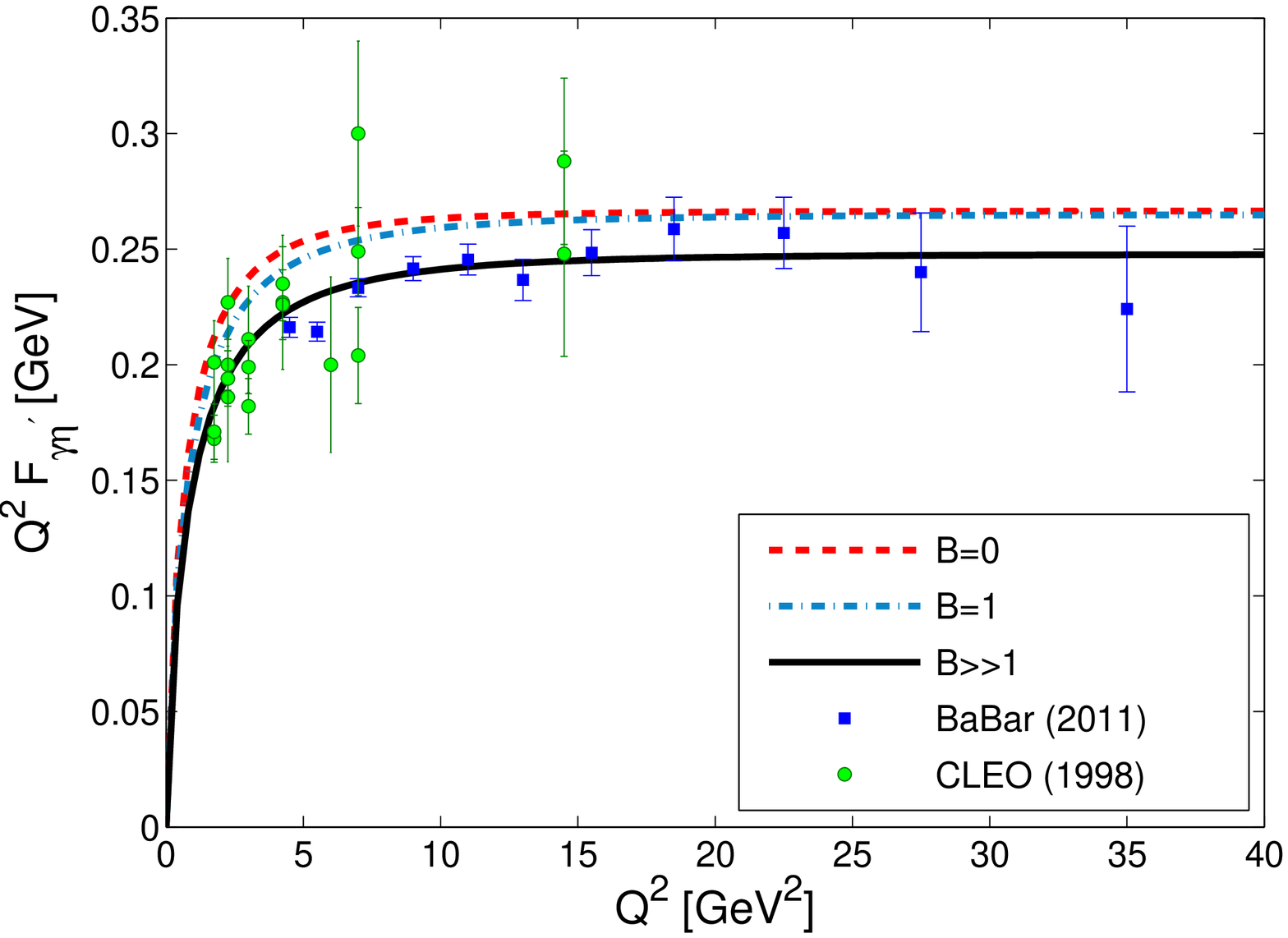}
\caption{Our predictions for the $\eta$ and $\eta^\prime$ transition form factors compared to the data from the BaBar \cite{BABAR:2011ad} and CLEO \cite{Gronberg:1997fj} Collaborations. For these predictions, we have set $B_q=B_s=B$. Dashed red curves: $B=0$. Dotted-dashed blue curves: $B=1$. Solid black curves: $B \gg 1$.} 
\label{Fig:eta-etaprime-TFF}
\end{figure}

Our predictions are shown in Fig. \ref{Fig:eta-etaprime-TFF}. As can be seen, our predictions for the $\eta$ and $\eta^\prime$ agree very well with the data when dynamical spin effects are maximal in both mesons. For the $\eta^\prime$, this is in agreement with our earlier findings regarding the radiative decay width but for the $\eta$, it means that there is a tension between the spacelike transition form factor data and the measured radiative decay width with respect to the importance of dynamical spin effects. The radiative decay width is sensitive to the transition form factor $Q^2=0$: $F_{\eta \gamma}(0)$ which we computed using the ABJ anomaly relations. It would be useful to compute this quantity without recourse to the ABJ anomaly relations and instead by finding the simultaneous $Q^2 \to 0$ limits of the  spacelike and timelike transition form factors. This requires the determination of the timelike transition form factor either by analytic continuation of the spacelike form factor or otherwise \cite{Choi:2017zxn}. We do not carry out this task in this paper but simply state that it is perhaps more likely that dynamical spin effects are also maximal in the $\eta$ as revealed by the spacelike transition form factor data which span a wider kinematical range: $4~\mathrm{GeV}^2 < Q^2 < 40~\mathrm{GeV}^2$ than the single radiative decay width datum which is sensitive to the transition form factor at $Q^2=0$. However, in our present analysis, we have completely ignored the possible effects of gluonic or charm components in the $\eta^\prime$  \cite{Feldmann:1998sh,Feldmann:1998vh,Kou:1999tt,Harland-Lang:2013ncy,Harland-Lang:2017mse,Thomas:2007uy,Ambrosino:2009sc,Ambrosino:2006gk,Ahmady:1998mi}. A fuller analysis is required to arrive at a definite conclusion.

Interestingly, we also find that our predictions for the spacelike transition form factors of both $\eta$ and $\eta^\prime$ hardly change if we restrict maximal dynamical spin effects only to their nonstrange $q\bar{q}$ sector. In other words, it does not matter much if we take $[B_q \gg 1, B_s \gg 1]$ or $[B_q \gg 1, B_s =0]$. This observation is consistent with our earlier findings that the pion data favor maximal dynamical spin effects and that the kaon data prefer no such effects. Maximal dynamical spin effects thus seem to be a signature of the nonstrange sector $q\bar{q}$ in the pseudoscalar mesons while such effects are likely to be suppressed by the presence of the strange quark ($\bar{q}s$ or $\bar{s} s$).

\section{Quark spins in the pion}
As can be seen from Eq. \eqref{spin-improved-wfn-b}, dynamical spin effects lead to the possibility that the quark and antiquark have aligned spins in the pseudoscalar meson. Since we are now confident that such effects are maximal in the pion, it is instructive to compute the probability to find its valence quark and antiquark with aligned spins. Recall that in the original semiclassical approximation of light-front holographic QCD, this probability is zero by assumption. 

\begin{figure}[htbp]
\centering 
\includegraphics[width=10cm,height=10cm]{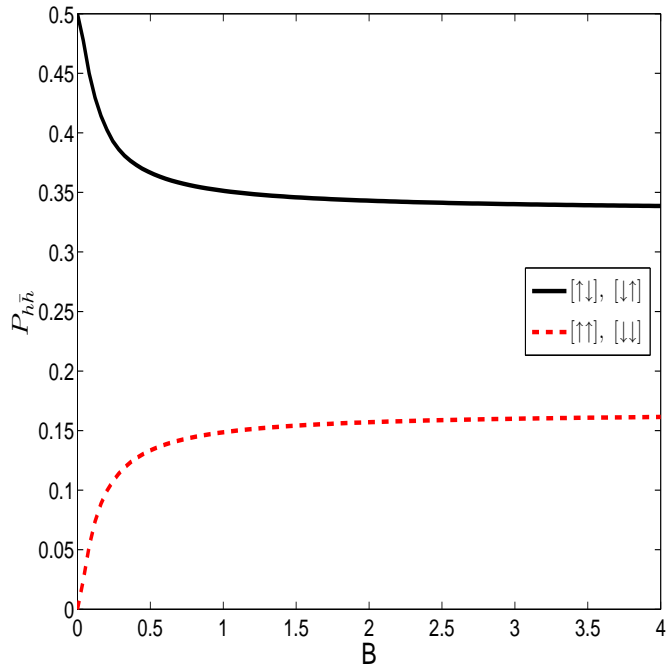}
\caption{The probabilities of finding aligned or opposite spins in the pion as a function of the dynamical spin parameter $B$.} 
\label{Fig:Probability}
\end{figure}

\begin{figure}[htbp]
\centering 
\includegraphics[width=8cm,height=8cm]{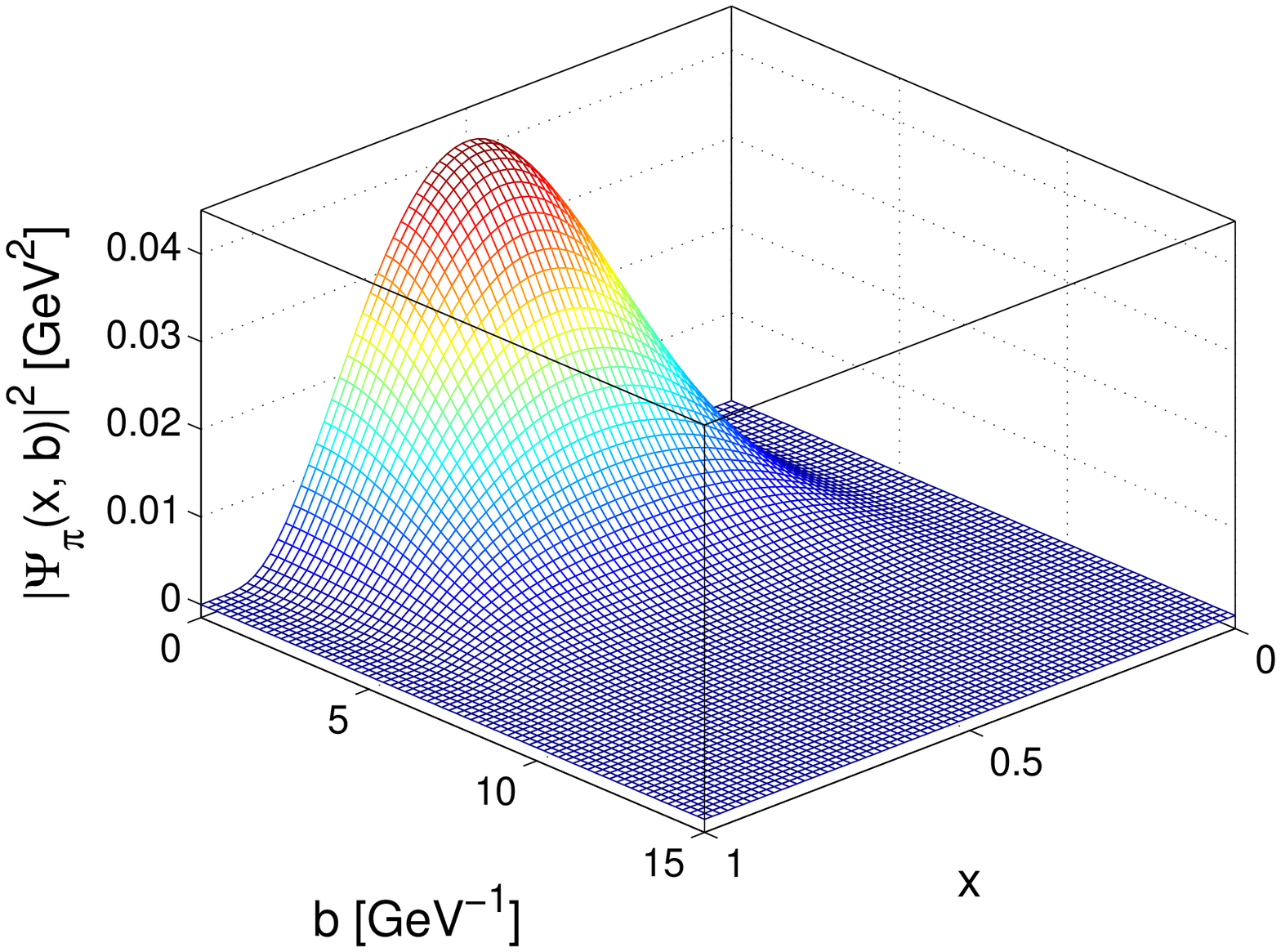}\includegraphics[width=8cm,height=8cm]{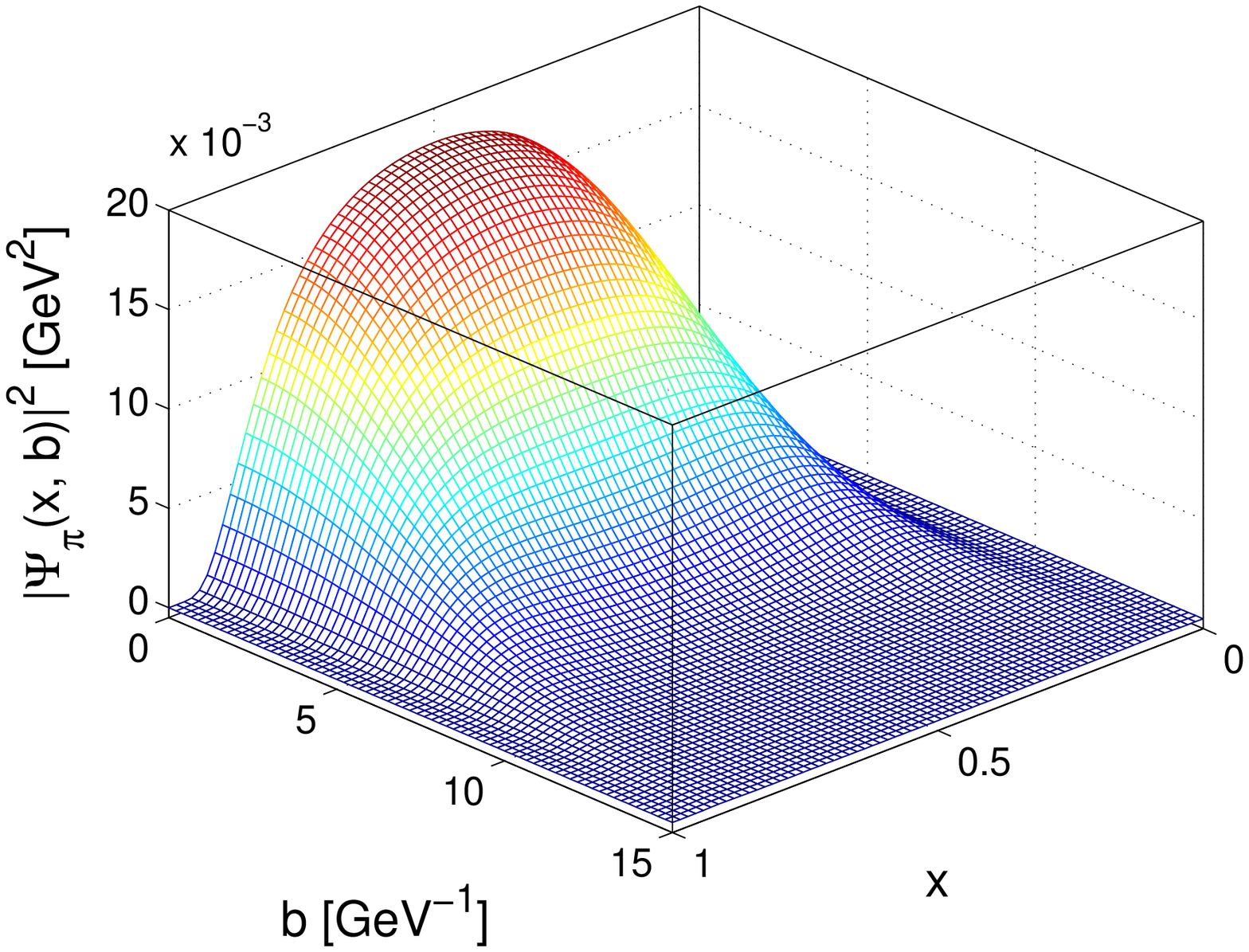}
\caption{The holographic pion light-front wavefunction squared with $B=0$ (left) and with $B \gg 1$ (right). } 
\label{Fig:Pionwf}
\end{figure}

Expanding the pion state in helicity space, we have

\begin{equation}
	|\pi \rangle=\sum_{h\bar{h}} P_{h\bar{h}} |h,\bar{h} \rangle
	\end{equation}
with
\begin{equation}
	P_{h\bar{h}}= \int \mathrm{d}^2 \mathbf{b}  \mathrm{d} x |\Psi^{\pi}_{h,\bar{h}}(x,\mathbf{b})|^2 
\end{equation}	
where $\Psi^{\pi}_{h,\bar{h}}(x,\mathbf{b})$ is given by Eq. \eqref{spin-improved-wfn-b}. $P_{h\bar{h}}$ is the probability to find the quark and antiquark in the helicity configuration $[h,\bar{h}]$ in the meson. The possibility  that both spin configurations (aligned and opposite) should be present in the pion was  suggested long ago in Ref. \cite{Leutwyler:1973mu,Leutwyler:1973tn}. In Fig. \ref{Fig:Probability}, we show the probabilities of finding aligned or opposite spins in the pion. As can be seen, the probability of finding aligned spins is zero for $B=0$ (no dynamical spin) and it saturates to $15\%$ for $B \gg 1$. Therefore, we predict that the probability that the quark and antiquark have aligned spins ($[h \bar{h}]=[\uparrow \uparrow]~\text{or}~[\downarrow \downarrow]$) in the pion is $30\%$. Thus, finding that the quark and antiquark have antialigned is still more probable ($70\%$). Finally, we illustrate dynamical spin effects in the holographic pion wavefunction after squaring and summing over all quark helicities in Fig. \ref{Fig:Pionwf}. As can be seen, the dynamical spin effects significantly broaden the wavefunction.

\section{Conclusions}
We have shown that dynamical spin effects are maximal in the pion holographic light-front wavefunction, where they lead to a successful simultaneous description of a wide range of data on the decay constant, radiative decay width, charge radius, EM and transition form factors and also, after QCD evolution, the Fermilab PDF and PDA data. We predict up to a $30\%$ probability that the spins of its valence quark and antiquark are aligned. On the other hand, the smaller kaon data set (decay constant, charge radius, EM form factor) prefers no dynamical spin effects. For the $\eta{-}\eta^\prime$ system, the measured radiative widths reveal dynamical spin effects only in $\eta^\prime$ but the data on their transition form factors clearly prefer dynamical spin effects in both mesons, even if such effects are restricted to their nonstrange sectors. We conclude that the importance of dynamical spin effects is not so much correlated to the mass of the pseudoscalar meson but instead to its quark flavor content.

\section{Acknowledgements}
M.A. and R.S. are supported by individual Discovery Grants from the National Science and Engineering Research Council of Canada (NSERC): No. SAPIN-2017-00033 and No. SAPIN-2017-00031, respectively. C.M. is supported by  the China Postdoctoral Science Foundation under the Grant No. 2017M623279. We thank William James Woodley for his input in coding.

\bibliographystyle{apsrev}
\bibliography{Pion2018_final.bib}

\end{document}